\documentclass[aps,prd,twocolumn,groupedaddress,showpacs,showkeys]{revtex4-1} 

\usepackage[utf8]{inputenc} 

\usepackage{enumitem}

\usepackage{graphicx}
\usepackage{amssymb}
\usepackage{amsmath}
\usepackage{color}
\usepackage{todonotes}
\usepackage[normalem]{ulem}
\definecolor{hartmutdirkcorrect}{rgb}{.7,.7,.2}
\definecolor{dirkcorrect}{rgb}{.7,.2,.2}
\definecolor{dirksuggest}{rgb}{.2,.2,.7}
\newcommand{\dcor}{\color{dirkcorrect}}

\newcommand{\black}{\color{black}}

\begin{document}
\title{On Radiation Reaction in Classical Electrodynamics}
\author{C.\ Bild}
\affiliation{LMU, ASC, Munich, Germany}
\email[]{christian.bild@physik.uni-muenchen.de}
\email[]{hartmut.ruhl@physik.uni-muenchen.de}
\author{H.\ Ruhl}
\affiliation{LMU, ASC, Munich, Germany}
\author{D.-A.\ Deckert}
\affiliation{Mathematical Institute, LMU Munich, Germany}
\email[]{dirk.deckert@math.lmu.de}

\date{\today}

\begin{abstract} 
    The Lorentz-Abraham-Dirac equations (LAD)  may be  the most commonly
    accepted equation  describing the motion of  a classical charged particle
    in its electromagnetic field. However, it is well known that they bear several
    problems.  In particular, almost all solutions are dynamically unstable,
    and therefore,  highly questionable.  As shown by Spohn et al., stable
    solutions to LAD can be approximated by means of singular
    perturbation theory in a certain regime and lead to the Landau-Lifshitz
    equation (LL).  However, for two charges there are also
    counter-examples,
    in which all solutions to LAD are
    unstable.  The question remains whether better equations of motion
    than LAD can be found to describe the dynamics of charges in the
    electromagnetic fields. In this paper we present an approach to derive such
    equations of motions, taking as input the Maxwell equations and a
    particular charge model only, similar to the model suggested by Dirac in his original
    derivation of LAD in 1938. We present a candidate for new equations
    of motion for the case of a single charge. Our approach is
    motivated by the observation that Dirac's
    derivation relies on an unjustified application of Stokes'
    theorem and an equally unjustified Taylor expansion of terms in his
    evolution equations. For this purpose, Dirac's calculation is repeated using
    an extended charge model that does allow for the application of
    Stokes' theorem and enables us to find an explicit equation of motion by
    adapting Parrott's derivation,
    thus avoiding a Taylor expansion. The result are second order differential
    delay equations, which describe the radiation reaction force for the charge
    model at hand. Their informal Taylor expansion in the radius of the charge
    model used in the paper reveals again the famous triple dot term
    of LAD but provokes the mentioned dynamical instability by a mechanism we
    discuss and, as the derived equations of motion are explicit, 
    is unnecessary.
\end{abstract}

\pacs{03.50.DE, 41.60.-m}
\keywords{radiation reaction, delay equations, classical electrodynamics}

\maketitle   

\section{The Lorentz-Abraham-Dirac equation}
Finding an equation of motion for a classical charged particle in its classical
radiation field is a very old problem; see the exhaustive references in
\cite{rohrlich2007classical,lyle2010self,spohn2004dynamics,penrose1990emperor}.
Since point particles lead to divergences within classical electrodynamics,
different remedies have been explored.
One approach is to modify Maxwell's
equations as has been done by Born-Infeld \cite{born1934foundations} or
Bopp-Podolsky \cite{podolsky1948review}, see also
\cite{kiessling2004electromagnetic, kiessling2019force} and other
approaches such as \cite{feynman1948relativistic} and more recently
\cite{raju2011radiative}.
Another approach is to
introduce an extended charge model as it has been done by Abraham
\cite{St.M.1906}, Lorentz \cite{lorentz2007theory}, and many others
\cite{caldirola1956new}. Beside their tension with regard to Lorentz
invariance, very early it was realized that such models introduce an
electrodynamic inertial mass for which Dirac proposed his famous mass
renormalization program to investigate the corresponding point charge limit
\cite{dirac1938classical}; see \cite{PhysRevD.80.024031} for a recent approach
in controlling such a point-charge limit. An entirely different approach was
taken by Wheeler and Feynman \cite{wheeler1945interaction} who were able to
derive an radiation reaction equation from an action-at-a-distance principle at
the cost of introducing advanced and retarded delays in the equations of
motion. Beside the problem of self-interaction, it is interesting to note that
in the case of more than one interacting point-charges there are further
difficulties connected to the emergence of singular fronts in the solutions to
the Maxwell equations \cite{hartenstein2016}. 

Although all these approaches are quite different, the
Lorentz-Abraham-Dirac (LAD) equations of motion almost always
appear as a limiting case. Hence, whatever the fundamental equations
of motion for a classical charged particle in its radiation field are,
the general consensus would likely be that a connection to the LAD
equations should be possible in a certain limit. At this point it is
interesting to note, as pointed out in \cite{spohn2004dynamics}, that
there is no experiment that could measure the radiative corrections to
the corresponding charge trajectories introduced by any of the
candidates of radiation reaction equations with sufficient precision even
though, in a large regime, the phenomenon of radiation reaction is a
purely classical effect. However, recently radiation reaction has
attracted new attention
\cite{PhysRevX.8.011020,sheffer2018towards,wistisen2018experimental},
  which gives hope
that accurate experimental data will be provided in the future. The
LAD equations are given by 
\begin{equation}\label{ald}
    m a^\alpha=q F^{\alpha\beta}(z) \, u_\beta +\frac{2q^2}{3}
\left( \frac{d a}{d\tau}^\alpha +a^\beta a_\beta u^\alpha \right) \, ,
\end{equation}
where $z^\alpha(\tau)$, $u^\alpha(\tau)$, and $a^\alpha(\tau)$ denote
the relativistic position, velocity, and acceleration four-vectors of
the charge under examination, respectively, with $\tau$ being the
world-line parameter, e.g., the proper time. Moreover, $m$ denotes its
effective inertial mass, $q$ its charge, and $F^{\alpha\beta}$ is the
field strength tensor of the electromagnetic fields of all other
particles that may also include an additional external field. 
Throughout the paper we set the speed of light to $c=1$.
Hence, the first expression on
the right-hand side of \eqref{ald} is the Lorentz force due to all other charges and the external field.
The second expression on the right-hand side
describes the so-called self-interaction, i.e., the interaction of the charge
under consideration with its own radiation field. Since this term involves a
third derivative of the world-line
$z^\alpha(\tau)$ one also refers to it as radiation friction term.
There is no straight forward
way to arrive at expression (\ref{ald}). In Dirac's paper
\cite{dirac1938classical} it is the zero order term of the total self-force, i.e., 
the Lorentz force on the charge through its own Maxwell field,
expanded in a Taylor series about the radius $\epsilon$ of the charge
distribution. In Dirac computation, there is also a term of order
$\epsilon^{-1}$. This
term is proportional to the acceleration. It is usually brought to the left
side of equation (\ref{ald}) and absorbed in the mass coefficient such
that
\begin{equation}
    \label{eq:renorm}
    \dcor
    \black m_{\text{ren}}=m+\frac{q^2}{2\epsilon} \, ,
\end{equation}
where $m$ is the bare inertial mass of the charged
particle and $m_{\text{ren}}$ the renormalized one.
The usual argument in the text books is that the bare inertial mass and the
inertial mass originating from the
field energy, cannot be separated by any experiment and only their sum can be
observed. While this is surely a sensible argument, it has to be
emphasized that for $\epsilon$ smaller than the classical electron radius
$e^2/(4 \pi
\epsilon_0 m_e c^2)$ the argument implies that the bare mass $m$ has to
be negative in order to ensure that
the electron attains the inertial mass known from experiments. It has
been emphasized that this implication even holds true for any extended
charge model and is not just an artifact of the limit $\epsilon\to
0$. 

Although this renormalization procedure has been the
reason for some concern it seems to be unavoidable if one is not willing to
modify Maxwell equations or the Lorentz force and still wants to describe a
relativistic particle as light and small as the electron seems to
be. It is also important to note that there is no easy way out, e.g.,
by claiming that on such scales quantum electrodynamics (QED) would have to be
invoked to describe the phenomenon of radiation reaction. First, QED has been
plagued by exactly the same problem of infinities through self-interaction --
there, called the ultraviolet divergence of the photon field, which
has prevented the formulation of a mathematically well-defined Schrödinger-type equation for the
dynamics ever since. And second, in a large regime the quantum corrections do
not seem to play an important role. For ultra-strong electromagnetic
backgrounds, however, observable signatures of the nonlinear quantum
vacuum as well as a subtle interesting interplay with radiation reaction 
are to be expected \cite{king2014interaction}. Due to recent progress
in technology (CALA, ELI) the correct formulation of both the classical
and quantum dynamics of radiation reaction has regained high priority.

All higher order terms in $\epsilon$ in Dirac's computation
depend on assumptions about the geometry of the current distribution and
usually are neglected by taking the limit $\epsilon\to 0$. By all means, it
is justified to worry if taking the limit
$\epsilon\rightarrow 0$ leads to a well-behaved equation of motion. Foremost, this limit is taken
at a fixed instant in time only. However, to control the difference of potential solutions for
varying $\epsilon$, bounds at least uniform on a time interval are required.
Dirac himself
pointed out that even for the case of a single particle in the absence of
external fields there is but one physical sensible LAD solution, namely the
straight line, while all other solutions describe charges that
accelerate increasingly in time.

An example of how neglecting higher order terms in a Taylor series
can lead to unstable solutions is given in chapter \ref{taylor} of
this paper. One example for such a solution of (\ref{ald}) is
\begin{equation}\label{fourvel}
u^\alpha(\tau)=\begin{pmatrix}
\cosh\left(e^{\frac{3m}{2q^2}\tau}\right)\\0\\0\\\sinh\left(e^{\frac{3m}{2q^2}\tau}\right)
\end{pmatrix}
\, ,
\end{equation}
which are obviously highly questionable. They
are referred to as run-away solutions. Believing in the
physical relevance of the LAD equations implies finding a way to
rule out run-away solutions.
Since the LAD equations are third order equations, the initial value
problem admits points from a nine-dimensional manifold, i.e., position,
momentum, and acceleration three-vectors at one time instant.  One
approach is using singular perturbation theory
in the leading part of the second term of \eqref{ald} in the approximation
of slowly varying external fields,
which results in the Landau-Lifschitz (LL) equation; see
\cite{spohn2004dynamics} for an extensive overview. In the perturbative regime and
for the case of a single charge it can be shown that all stable solutions have initial
values on a six-dimensional sub-manifold, i.e., comprising position and momentum
three-vectors at an instant of time only from which the ``correct'' initial
acceleration can in principle be computed. The stable solutions are the
ones that are approximates by the LL equations; see \cite{di2008exact} for an exact
solution. 
The LL equations are therefore dynamically well behaved
and also useful for practical calculations in there range of
validity. Strictly speaking, however, they are in character more an approximation rather than a 
fundamental equations. The strategy to simply select the ``correct'' initial
acceleration fails in more complicated systems. This is 
shown by Eliezer \cite{eliezer1943hydrogen} by giving a 
counter example. Eliezer considers two oppositely charged particles moving
towards each other in a symmetric fashion and proves that
for {\em all} initial accelerations, the particles turn around at some point
before they collide and fly apart with ever increasing
acceleration. His result implies that there exist cases in which the LAD
equations do not seem to give a satisfactory answer. The example by Eliezer is
elaborately discussed by Parrott in \cite{parrott2012relativistic}. 
At the very least for those cases, new equations of motion are needed, but also
in general, having access only to stable approximate solutions does not seem to be entirely
satisfactory.

This present unsatisfactory situation is the main motivation for our
work. We will reconsider Dirac's and Parrot's derivations of 
radiation reaction equations and by adapting and extending them
propose new exact equations of motion, i.e., without making use of a
Taylor expansion.

\subsection{\label{dirac}Dirac's original derivation}
\begin{figure}
	\includegraphics[height=5cm]{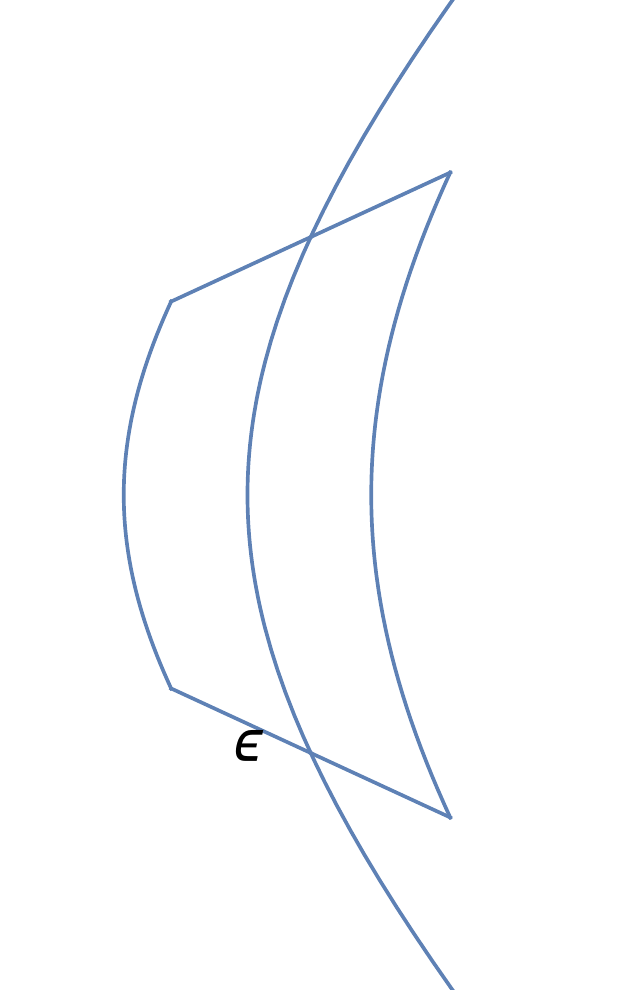}
	\caption{\label{tubedirac} Dirac's tube, it is defined by
          moving the length $\epsilon$ along the hyperplane of
          simultaneity in the rest frame in all directions between
          $\tau_1$ and $\tau_2$.}
\end{figure}

To obtain ``better'' equations of motion, as compared to
LAD, it is important to understand the shortcomings in their derivation.
Dirac makes use of a point particle as the model of
a charged particle. His approach has the advantage that he does not need to be
concerned about the inner structure of the particle. The
disadvantage, however, is that the Lorentz force
cannot be used right away
because the fields are singular in the vicinity of
the point charge. Instead of using the Lorentz force to infer the
equation of motion, Dirac uses the concept of energy-momentum
conservation as a starting point since the change in momentum of the charge can
be expressed by means of energy-momentum tensor
\begin{eqnarray}
\label{energy-momentum-tensor}
&&4 \pi T^{\alpha\beta} = F^{\alpha\gamma} F_\gamma^\beta +
    \frac{1}{4}\eta^{\alpha\beta} F^{\gamma\delta} F_{\gamma\delta} \, .
\end{eqnarray}
In (\ref{energy-momentum-tensor}) the quantity $\eta^{\alpha \beta}$ is the metric tensor 
having the signature $\eta=\operatorname{diag}(1,-1,-1,-1)$. 
Now let $V(\tau_1,\tau_2)$ be a smooth space-time region, which
encompasses an interval of the world-line of the charge given by
$z^\alpha$ with the entry and exit space-time points
$z^\alpha(\tau_1)$ and $z^\alpha(\tau_2)$, respectively. Dirac
implicitly argues in the spirit of Stokes' theorem that the
volume integral over $V(\tau_1,\tau_2)$ of the divergence of
$T^{\alpha\beta}$ equals the surface integral over the boundary
$\partial V(\tau_1,\tau_2)$ of the energy-momentum flow out of the
volume. Thus, we obtain
\begin{eqnarray}
\label{gaus}
&&P^{\alpha}(\tau_2)-P^{\alpha}(\tau_1)=\int_{\tau_1}^{\tau_2} d\tau \,
F^\alpha(z(\tau)) \nonumber \\
&&=\int_{\tau_1}^{\tau_2} d\tau \, q
F^{\alpha\beta}(x) u_\beta(\tau)\nonumber\\
&&=\int_{ V(\tau_1,\tau_2) } d^4x \int   d\tau \, q
F^{\alpha\beta}(x)  \, u_\beta(\tau) \,
\delta^4(z^\beta(\tau)-x^\beta)\nonumber\\
&&=\int_{ V(\tau_1,\tau_2) } d^4x \, F^{\alpha\beta}(x) \,
j_\beta(x) \nonumber \\
&&= - \int_{ V(\tau_1,\tau_2) } d^4x \,
\partial_\beta T^{\alpha\beta}(x)\nonumber\\
&&=-\int_{ \partial V(\tau_1,\tau_2) } d^3x_\beta \,
   T^{\alpha\beta}(x) \, ,
\end{eqnarray}
where the difference $P^{\alpha}(\tau_2) - P^{\alpha}(\tau_1)$ in
(\ref{gaus}) is the total change of momentum of the point charge along
the world-line $z^{\alpha}(\tau)$. The surface measure times the
normal four-vector $n^\beta(x)$ on the boundary $\partial
V(\tau_1,\tau_2)$ is denoted by $d^3x^\beta$. In (\ref{gaus})  use
has been made of the definition of the current density of a point
particle 
\begin{eqnarray}
    &&j^\alpha \left( x\right)=q\int d\tau \, u^\alpha (\tau) \,
       \delta^4 \left( z^\beta(\tau) - x^\beta \right) \, . 
\end{eqnarray}
Unfortunately, Stokes' theorem is not applicable in the context of the
assumptions made by Dirac as the fields
$F^{\alpha\beta}$ that enter $T^{\alpha\beta}$ are not smooth
but singular on $V(\tau_1,\tau_2)$ due to the point
charge model. As a matter of fact, neither the left- nor
right-hand side of equation \eqref{gaus} is well-defined.
In the expressions on the right-hand side, however,
the field divergences appear only at the points where the
particle enters and leaves the integration volume $V(\tau_1,\tau_2)$.
In order to treat the integrations there Dirac introduced a cut-off to remove the
divergent contributions. The definition and physical meaning of a
cut-off is discussed in chapter \ref{cut-off}. Dirac argues that the
shape of the integration volume does not influence the final result
since the divergence of the energy-momentum tensor vanishes at points with no
charge present. Hence, only the amount of the charge inside the volume matters
and not its shape. However, we will see that this is not
true for the point-charge model assumed by Dirac and that the shape of
the volume actually matters at the points where the world-line 
penetrates the surface of the volume. Next, Dirac picks as the volume
$V(\tau_1,\tau_2)$ a four dimensional tube
consisting of the union of spheres with radii of the
size of the cut-off parameter $\epsilon$ in each rest frame
between the two fixed entry and exit space-time points at 
$z^\alpha(\tau_1)$ and $z^\alpha(\tau_2)$. Dirac's tube is visualized
in Fig.~\ref{tubedirac}. Dirac divides the surface integration into
two parts, an integration over the lateral surface of his tube and an
integration over the caps. While Dirac presents an explicit
calculation of the contribution of the lateral surface to the
energy-momentum tensor, he is not performing the cap
integrations, which would diverge for the point particle. Instead, he
guesses that the cap integrals are equivalent to the kinetic term
$ma^\alpha$. Dirac's guess in fact implies a cut-off in the fields
since the contribution of the cap integrals to the
energy-momentum tensor is assumed to be zero.
The remaining integral over the lateral surface of the tube is always
close to the world-line. For the evaluation of the fields at the
lateral surface of the tube Dirac needs the retarded proper time. An explicit
expression of the latter, however, is generally not available. Hence,
Dirac introduces a Taylor series in the cut-off parameter $\epsilon$
and assumes that all higher order terms of the latter only give
negligible contributions to the dynamics provided the cut-off parameter
is small enough. Dirac's assumption, however, is unjustified as we will
discuss later by means of a counter example. By differentiation of
(\ref{gaus}) with respect to $\tau_2$ Dirac obtains an expression for
the Lorentz force at time $\tau_2$. To calculate the surface integrals
implied in (\ref{gaus}), Dirac determines the
corresponding Liénard-Wiechert potentials and calculates the
field-strength tensor. Finally, he computes the
energy-momentum tensor and carries out the integrations as discussed.
After all these steps and absorbing terms of the order $\epsilon^{-1}$
into the bare inertial mass according to \eqref {eq:renorm}, he
arrives at (\ref{ald}).

These issues and how to circumvent them will be the content of the next
sections. The outline of the paper is as follows. In chapter \ref{cut-off} it
is
discussed that the assumption of a cut-off and the requirement of
consistency with the Maxwell equations
imply an extended charge model. In chapter \ref{taylor} it is shown
that the Taylor series mentioned before cannot be used. In chapter
\ref{capform} the approach pursued by Parrott is discussed, which
allows to avoid the Taylor series. In the same
chapter a constraint that seems to be missing in Parrott's calculation on the tube
geometry is also discussed, which arises from the fact that the
total self-force on the particle is given by integration over the
Lorentz force density acting on the extended particle. 
This leads to the conclusion that the caps of the tube have to be
hyperplanes of simultaneity in the co-moving reference frame of the
charge. In chapter \ref{mainresult} an expression
for the radiation reaction force is derived, which is the first main result
of this paper. In chapter \ref{eff-eqn} new equations of motion
and a discussion of the resulting radiation reaction force are
given which represents the second main result of this work.

\subsection{\label{cut-off}Interpretation of the cut-off}
There is no obvious reason why the cap integrals of the energy-momentum
tensor appearing in (\ref{gaus}) at $\tau_1$ and $\tau_2$ with radius
$\epsilon$ can be neglected. No matter how small $\epsilon$ is 
the corresponding integrals give infinite contributions, which in view
of Stokes' theorem also depend on the geometry of the corresponding
cut-off (for $\epsilon \to 0$ also on the mode of convergence) and
therefore cannot be ignored.  In Dirac's derivation the cap
contributions are dropped, nevertheless. However, it is 
possible to give a reasonable interpretation of Dirac's cut-off even
without taking the limit. We note that the cap integrations at
$\tau_1$ and $\tau_2$ correspond to integrations over spheres at
$\tau_1$ and $\tau_2$. Obviously, the integrals over a sphere with
radius $\epsilon$ can be ignored if and only if the value of the sum
of the integrands for the spheres at $\tau_1$ and $\tau_2$ is
zero. This is not the case for a point particle but it is certainly
the case for a specific class of charge current distributions. The simplest
example of such a distribution is one which has no fields
inside of such a sphere. Thus, dropping the cap integrals in (\ref{gaus})
implies that the original field strength tensor of a point charge is
replaced by a field strength tensor which is zero inside a cut-off
region and identical to the field strength tensor of the point
particle outside of it. The corresponding distribution can be
calculated with the help of Maxwell's equations
\begin{equation}\label{j}
\partial_\alpha F^{\alpha\beta}_{\epsilon}=4 \pi j_{\epsilon}^\beta \, ,
\end{equation}
where $j^\beta_{\epsilon}$ is the new distribution due to the cut-off
$\epsilon$. Stokes' theorem shows that this distribution is located on the
surface of the sphere. But it is not necessarily homogeneous and, hence, does
not imply that the introduction of such a cut-off is nothing else than
replacing the original point charge by an extended current distribution on a
sphere and that taking the limit $\epsilon \to 0$ means shrinking the radius of
the distribution down to zero. In contrast, the general situation is more
subtle as even the limit $\epsilon\to 0$ involves
a choice, i.e., the mode of convergence of the particularly chosen
current model to the point-charge limit.

Throughout this work we will, however, keep $\epsilon>0$. Since the field
strength tensor of such a current distribution is free of divergences, Stokes'
theorem can be applied in the argument in (\ref{gaus}).

\subsection{\label{taylor} Taylor expanding in the
  cut-off}
From the discussion in chapter \ref{cut-off} we conclude that an extended
current distribution has to be considered. We assume that the current
distribution is spherical with the cut-off radius $\epsilon>0$.  This choice
implies that the radiation reaction force will then involve a delay due to the
finite speed of light of the field propagating through the extended particle. This delay is a shared feature of all extended charge models as can be seen in \cite{sommerfeld1904simplified}, \cite{caldirola1956new} or \cite{levine1977motion}.

It is shown in this paper that the radiation reaction force indeed
leads to 2nd order delay-differential equations and that the third
order derivative $da^\alpha/d\tau$ in (\ref{ald}) originates from a
Taylor expansion in $\epsilon$ of the delayed radiation reaction
force. As it is well-known \cite{driver1977introduction,raju2011radiative,walther2014topics}, 
dropping all higher order terms in $\epsilon$ to obtain
(\ref{ald}) can lead to a severe change in the corresponding space of
solutions as can be demonstrated by the following simple 
example:
\begin{equation}
\label{simple-delay-eqn}
z(t)=z(t-\epsilon).
\end{equation}
The solutions to \eqref{simple-delay-eqn} are obviously periodic functions with period length
$\epsilon$. Taylor expanding informally the right-hand
side of (\ref{simple-delay-eqn}) up to 2nd order and truncating the rest gives 
\begin{equation}\label{Taylor}
z(t)=z(t)-\epsilon \dot{z}(t)+\frac{\epsilon^2}{2}\ddot{z}(t) \, .
\end{equation}
One solution of this equation is 
\begin{equation}\label{eq:example}
z(t)=e^\frac{2t}{\epsilon} \, .
\end{equation}
This is clearly no solution to the original equation \eqref{simple-delay-eqn}.
It exhibits a behavior much like the unstable solutions of the LAD equation,
the so-called run-away solutions.
The reason why a Taylor expansion of (\ref{simple-delay-eqn}) in
$\epsilon$ fails can be explained as follows. Although the right hand
sides of the two equations \eqref{simple-delay-eqn} and \eqref{Taylor} for comparable
initial conditions and at a fixed instant in time differ only by a term of
the order of $\epsilon^3$ the implication is not that also the two
respective solutions remain close to each other for other times. For
the latter one needs a uniform estimate of the difference of the
respective right-hand sides of \eqref{simple-delay-eqn} and \eqref{Taylor} on at least a
time interval, e.g., in the spirit of Grönwall's lemma. 

For our simple example we can readily compute the contribution coming
from the neglected higher-order terms. They are
\begin{equation}
\sum_{n=3}^{\infty} \frac{(-\epsilon)^n}{n!} z^{(n)} (t) = \left(
  e^{-2} -1 \right) \, z(t) \, .
\end{equation}
Thus, the smallness of higher-order terms does not directly depend
on $\epsilon$ but on the norm of the correspon\-ding solution $z(t)$. The latter
will in general depend on $\epsilon$ but in a much more subtle way. Controlling
it in $\epsilon$ therefore requires a careful mathematical
analysis. It is not sufficient to simply control the right-hand side of
\eqref{simple-delay-eqn} at one instant in time. The emergence of the run-away
solutions as \eqref{eq:example} after a Taylor expansion neglecting higher
orders in our simple example shows that higher order terms in
$\epsilon$ cannot be ignored in general.

The conclusion is that we have to repeat Dirac's calculation taking
the terms to all orders into account. This appears not to be feasible
for the tube Dirac has chosen. However, the calculation can be carried
out as outlined by Parrott \cite{parrott2012relativistic} for a tube
suggested by Bhabha. In chapter \ref{capform} the result of Parrott's
calculation and the need for modifications of the tube at the caps
used in our paper are discussed.

\subsection{\label{capform}Meaningful caps}
In his book Parrott \cite{parrott2012relativistic} repeats Dirac's
calculation without the Taylor expansion that Dirac uses. We argue
shortly why Parrott's calculation still has to be modified in order to
lead to a meaningful candidate for an equation of motion with
radiation damping. 

Parrott evaluates the time integral over the force in (\ref{gaus}),
which equals the time integral over the Lamor formula 
\begin{eqnarray}
\label{integral_lamor}
&& \int^{\tau_2}_{\tau_1} d\tau \,
F^\alpha(z(\tau)) \nonumber \\
&&= \int^{\tau_2}_{\tau_1} d\tau \, (2q^2/3) \, ( a^\beta
a_\beta \, u^\alpha ) (\tau) \, . 
\end{eqnarray}
Parrott does not carry out the time derivative of the expression
$\int^{\tau_2}_{\tau_1} d\tau \, F^\alpha(z(\tau))$ in
(\ref{integral_lamor}), which cannot be
computed for the tube used by Parrott. A valid force, as we argue, is however 
only obtained by performing the time derivative of
$\int^{\tau_2}_{\tau_1} d\tau \, F^\alpha(z(\tau))$. As a consequence,
Parrott's result may not be interpreted easily as a force, which also
manifests itself in the fact that the Lamor term is in general not
orthogonal to the four-velocity. Instead, Parrott argues that the
times $\tau_1$ and $\tau_2$ are somehow special. He requires that the
accelerations at $\tau_1$ and $\tau_2$ are zero. According to him, this
is a necessary condition if the result of the calculation must not
depend on the form of the caps. As a consequence, the time derivative
in Parrott's case is only possible for time regions with zero
acceleration, but for zero acceleration there is no radiation reaction
force. Since for $a^\alpha(\tau_1)=0$ and $a^\alpha(\tau_2)=0$ one
finds that
\begin{equation}
    \int^{\tau_2}_{\tau_1} d\tau \frac{2q^2}{3}
    \frac{d}{d\tau}a^\alpha(\tau)=0
\end{equation}
holds, Dirac's and Parrott's results agree when integrating Dirac's
force over time with the acceleration conditions above. Also Dirac's
result for the radiation reaction force depends on the choice of the
caps since Stokes' theorem cannot be applied the way Dirac argues as
we have outlined in chapter \ref{dirac}.

\begin{figure}
\includegraphics[width=4cm]{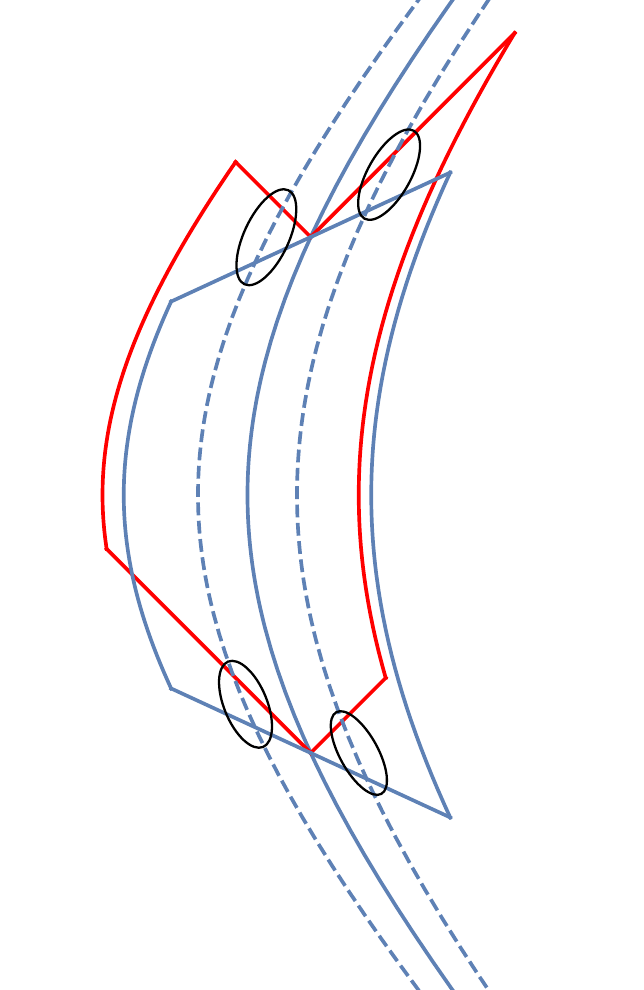}
\caption{\label{charge-amount-tubes} Due to the cut-off, the charge is distributed
around the world-line given by the blue solid line. The distributed charge is
represented by the dashed lines. Even though the red and the blue tubes start
and end at the same points of the world-line they contain different amounts of
charge as is highlighted by the circles. Note that the time axis is vertical
and the space axis horizontal. Also note that for any given definition
of the caps due to the theorem of Stokes the tube radius is irrelevant
as long as it is at least equal or larger than $\epsilon$. In the
paper we set the latter equal to $\epsilon$ contrary to
the sketch in the figure since then the cap contributions are
identical to zero.}
\end{figure}

The problem with Stokes' theorem can be illustrated nicely with the
help of an analogy. Let us consider the example of a point charge resting
at the origin of the coordinate system for which the fields are only
the Coulomb fields.
\begin{figure}
\includegraphics[height=3cm]{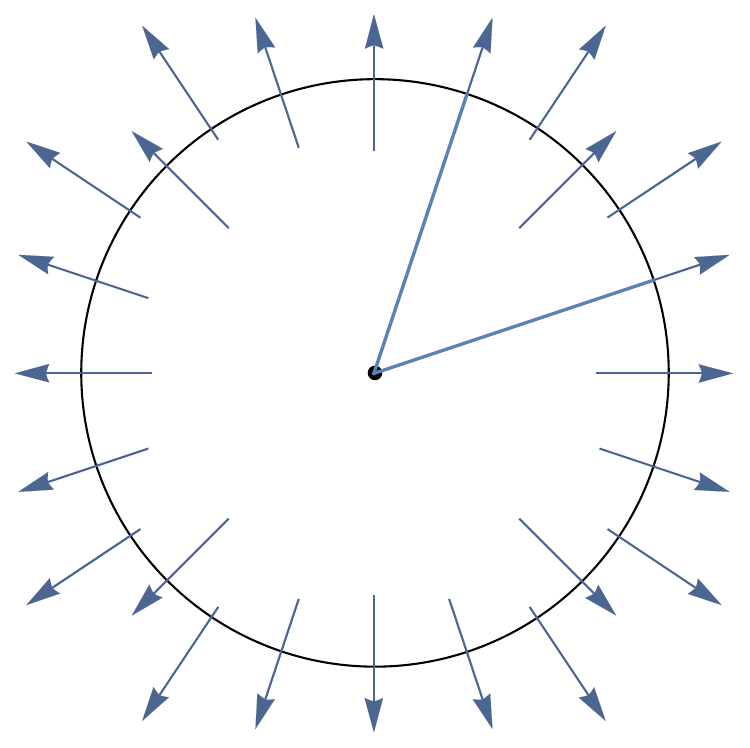}\qquad\qquad
\includegraphics[height=3cm]{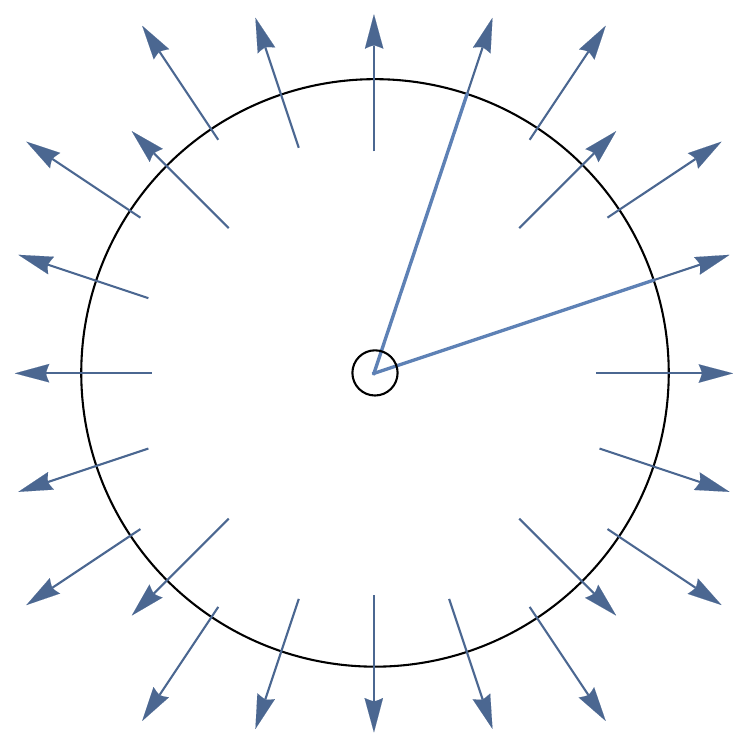}
\caption{\label{analogy}The left plot shows the flow of the Coulomb field of a resting point
charge through a sphere respectively a spherical sector. The right plot shows
the same situation for a charge corresponding to Coulomb fields with a cut-off
at radius $\epsilon$, which implies a charge model of a hollow
sphere of radius $\epsilon$ in the rest frame.}
\end{figure}
An integration of the flow of the electric field over the entire sphere around the origin
gives $4\pi \, q$, where $q$ is the charge at the origin. On the other
hand, an integration over a sector of the sphere gives $\Omega\, q$,
where $\Omega$ is the solid angle of the spherical sector. The lateral walls of the
spherical sector do not contribute since their normal vector is
orthogonal to the electric field. According to Stokes' theorem, as
used in Dirac's derivation, it is expected that volumes containing the
same amount of charge yield the same surface integrals of the flow of the
fields. Apparently, for a point charge on the surface the application of
Stokes' theorem does not yield unique results in contrast to what is
expected. To proceed with the analogy we cut off the field the way
Dirac does and as we have outlined in chapter \ref{cut-off}. According
to (\ref{j}) this implies that the point charge in its rest frame is
replaced by a homogeneously charged hollow
sphere with radius $\epsilon$. On its outside the hollow charge
distribution generates the same fields as a point charge while there
are no fields inside of it. For the hollow charge the integral over
the entire sphere yields the total charge
$4\pi \, q$ and the integral over a spherical sector the fraction
$\Omega \, q$ as before. In contrast to the situation of a point charge the
integration volumes now contain different amounts of charge in agreement
with Stokes' theorem as illustrated in Fig.~\ref{analogy}. Apparently,
the theorem of Stokes' can be applied after the introduction of the
cut-off. The implication is that the amount of charge contained in the
tubes depends on the choice of the caps as is illustrated with the
help of Fig.~\ref{charge-amount-tubes}.

Now we try to determine which amount of charge the tube should contain.
Since we are dealing with an extended current distribution, (\ref{gaus})
describes an integral over a force density which should be equal to
the momentum difference $P^\alpha(\tau_2)-P^\alpha(\tau_1)$, where
$P^\alpha(\tau)$ is the total momentum of the extended
particle. Performing the derivative of the force integral in
(\ref{gaus}) with respect to $\tau_2$ leads to
\begin{equation}\label{momentum-force}
\frac{dP^\alpha(\tau)}{d\tau}=F^\alpha(\tau) \, .
\end{equation} 
To obtain the correct total force $F^{\alpha}$ and total
momentum $P^{\alpha}$ in (\ref{momentum-force}) from the force and
momentum densities in (\ref{gaus}) the correct integration regions
have to be used. To obtain them and hence the correct tube geometry,
we consider the non-relativistic limit
\begin{equation}
	\vec{P}(t)=\int_{V} d^3r \, \vec{p} \left( t,\vec{r} \right) \, ,
\end{equation} 
where $\vec{p} \left( t,\vec{r} \right)$ is the momentum density. The
naive relativistic generalization 
\begin{equation}\label{naive}
 P^\alpha(\tau)=\int dx_1\wedge dx_2\wedge dx_3 \, p^\alpha (
   x^\beta )
\end{equation}
is not a Lorentz vector since the integration region, given by the
three form $dx_1\wedge dx_2\wedge dx_3$, is not a Lorentz scalar. The
reason for this is, that equal time surfaces get tilted under Lorentz
transformations. To find a relativistic generalization an
expression for the integration region is needed which is a Lorentz
scalar and reduces to (\ref{naive}) in the co-moving
coordinate frame in the non-relativistic limit. We consider the normal
vector of the integration region in equation (\ref{naive}). Formally
it can be obtained with the help of the Hodge dual and has the simple form
$(1,0,0,0)$ corresponding to $dt$. Since $u^\alpha dx_\alpha$ is a
Lorentz scalar, which reduces to $dt$ in the co-moving coordinate
frame, a good candidate for the integration region is given by the
Hodge dual of $-u^\alpha dx_\alpha$. This implies, that the
integration in (\ref{gaus}) should be performed over hyperplanes of
simultaneity in the rest frame. 

\begin{figure}
\includegraphics[width=8cm]{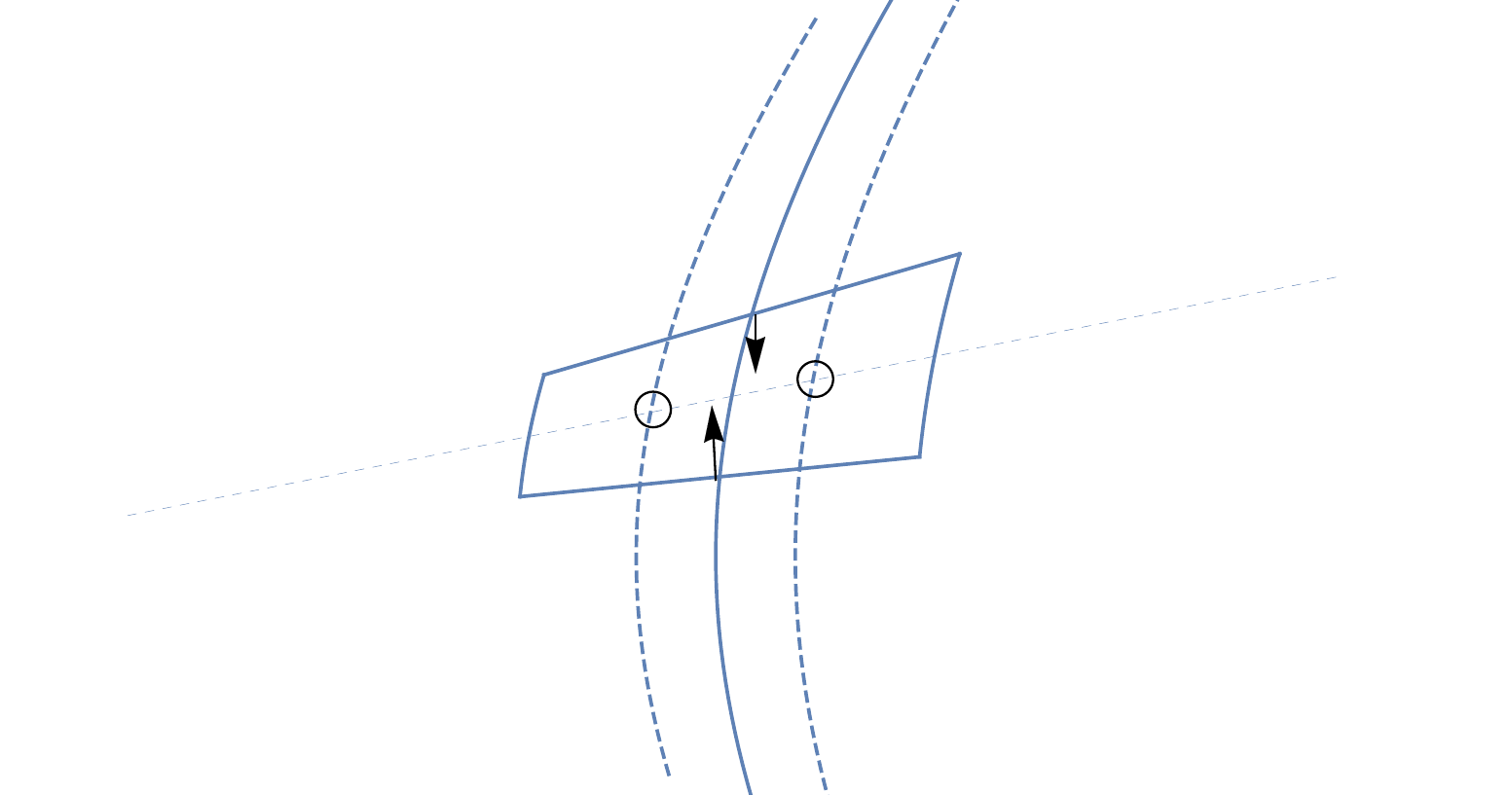}
\caption{\label{caps-of-tube} The current distribution is represented by the vertical
dashed lines. After taking the limit represented by the arrows only charge on
the hyperplane of simultaneity in the co-moving frame of reference represented
by the horizontal dashed line must be contained inside the tube.
This charge is marked by the two circles for better visibility. Hence,
the caps are hyperplanes of simultaneity in the co-moving reference
frame as can be seen from the plot.}
\end{figure}

The time derivative in equation (\ref{momentum-force}) can be interpreted
as the limit $\tau_1\rightarrow\tau_2$. Since in this limit the tube
is only allowed to contain charge located on a hyperplane of
simultaneity in the rest frame at time $\tau_2$, the caps also have to
be hyperplanes of simultaneity in their own rest frames, as can be
seen in Fig.~\ref{caps-of-tube}. By this line of reasoning we
conclude that Dirac's choice of the caps is the right one and
Parrott's extra condition $a^\alpha(\tau_i)=0$ is not needed for
Dirac's and our caps.

In chapter \ref{mainresult} we explain how to construct a tube in
such a way that both Parrott's approach as well as Dirac's caps can be used to
give a consistent derivation of a meaningful force. It is worth mentioning that
the results in the next chapters hold for any finite value of
$\epsilon$ and that the limit $\epsilon\rightarrow 0$ is never required
for explicit calculations.

\section{\label{mainresult}The radiation reaction force}
In this chapter we present the first main result of the paper, which in part is based
on Dirac's and Parrot's work but also goes beyond it by avoiding the
issues discussed in Section~\ref{dirac}. We provide new force candidate for the dynamics describing 
a charge in its radiation field. The corresponding equations of motions will be formulated and discussed
in the next Section~\ref{eff-eqn}. For this purpose we go back to Dirac's starting
point given in \eqref{gaus}, namely that the change in momentum of the charge
$P_\epsilon(\tau)$ can be inferred from the energy-momentum flow of its field
\begin{equation}
    \label{surface-integral-T}
    \partial_\tau P^{\alpha}_\epsilon \left( \tau \right)
    =
    \partial_\tau \int_{\partial V(\tau_1,\tau)}
    d^3x_\beta \, T_\epsilon^{\alpha\beta}(x) 
\end{equation}
Contrary to Dirac's consideration, we read this equation in terms of our charge
model defined by the $\epsilon$-dependent cut-off tube given in
Figure~\ref{tubenew}. To emphasize this difference, we add to all entities such
as the momentum and electromagnetic field derived from our charge model a
subscript $\epsilon$, while those derived from the point-charge model will not
carry this subscript.

The first goal is to compute the right-hand side of \eqref{surface-integral-T}.
This is carried out in Section~\ref{light-cone}-\ref{final-result}. The final
result is given \eqref{endgl} below. In order to infer a dynamical system that
couples the world-line $\tau\mapsto z^\alpha(\tau)$ to this computed 
momentum momentum in a self-consistent way, a relation between change of
momentum and change of velocity $\dot z^\alpha(\tau)$ has to be establish. This
final step is carried out in in Section~\ref{eff-eqn}.

\subsection{Light cone coordinates}\label{light-cone}
To carry out the calculation, the explicit shape of
$V(\tau_1,\tau)$, the expression for $T^{\alpha\beta}(x)$, and normal
vector $n_\beta(x)$, encoded in the surface measure $d^3x_\beta$, are
needed. Instead of the usual Cartesian coordinates $x^\alpha=(t, x_1,
x_2, x_3)$ it is more convenient to employ the so-called light-cone
coordinates $(\tau, r, \theta, \phi)$, which are introduced now. 
Given the space-time point $x^\alpha$ and a time-like world-line
$z^\alpha$ of the charge there exists a unique proper time $\tau$, such that
$z^\alpha(\tau)$ lies in the backward light-cone of $x^\alpha$, i.e., $\tau$ is
the unique solution of 
\begin{eqnarray}
&&\left( x^\alpha-z^\alpha(\tau) \right) \, \left( x_\alpha-z_\alpha(\tau) \right)=0
\end{eqnarray}
satisfying $x^0\geq z^0(\tau)$. The so-called retarded proper time
$\tau$ represents the first light-cone coordinate. The forward light
cone of $z^\alpha(\tau)$ can be viewed as consisting of spheres with
different radii. The radius $r$ of the sphere on which $x^\alpha$ lies
is the second light-cone coordinate. Since the distances in time and
in space of two points on the light cone for $c=1$ are equal, the
coordinate $r$ can be calculated by taking the zero component of
the four-vector $x^\alpha-z^\alpha(\tau)$ in the rest frame at the
retarded proper time $\tau$. Since the four-velocity of the charge
in the rest frame at the retarded proper time $\tau$ equals 
$u^\alpha(\tau)=(1, 0, 0, 0)$ we obtain
\begin{equation}
    r=\left( x^\alpha-z^\alpha(\tau) \right) \,
    u_\alpha(\tau) \, .
\end{equation}
To parametrize the points on the sphere in the rest frame defined by
$\tau$ and $r$ the spherical angles $\theta$ and $\phi$ are used, which
represent the third and fourth light-cone coordinates. 

The four-vector $x^\alpha-z^\alpha$ can now be split into space-like
and  time-like components
\begin{equation}
\label{split}
x^\alpha-z^\alpha(\tau)=r \,
\left(u^\alpha(\tau)+w^\alpha(\tau)\right) \, ,
\end{equation}
where the time-like component  in (\ref{split}) is given by the
four velocity $u^\alpha$ while the space-like component is
given by the four-vector $w^\alpha$, which is always space-like, of
length one, i.e. $w^\alpha \, w_\alpha=-1$, and orthogonal to the
four velocity, i.e. $w^\alpha \, u_\alpha=0$. In the rest frame
$w^\alpha(\tau)$ takes the form
\begin{equation}
    w^\alpha(\theta,\phi)=
\begin{pmatrix}
0\\ \sin \theta \, \cos \phi \, \\ \sin \theta \, \sin \phi \, \\ \cos \theta \,
\end{pmatrix}
\, .
\end{equation}
It is now possible to express $x^\alpha$ uniquely as a function of $\tau$, $r$,
$\theta$, and $\phi$. We obtain 
\begin{figure}
\center
\includegraphics[height=5cm]{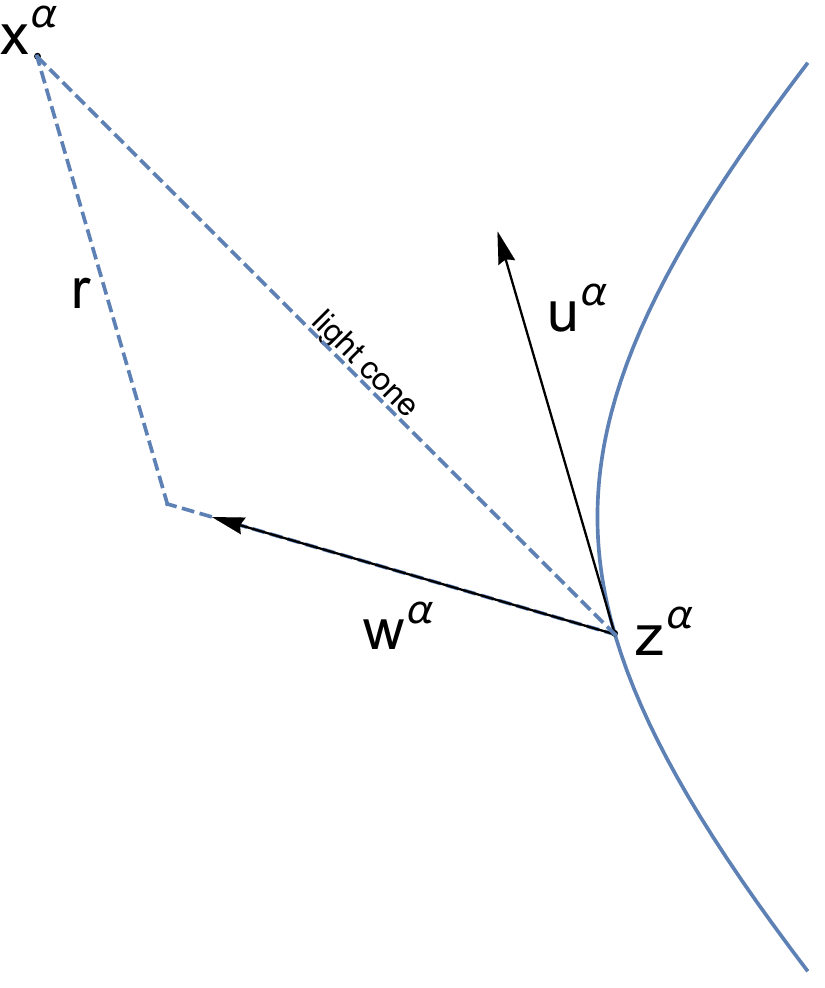}
\caption{\label{lightcone-plot} Representation of $z^\alpha,u^\alpha$, 
    $r$, and
$w^\alpha$.}
\end{figure}
\begin{equation}\label{lightcone}
    x^\alpha=z^\alpha(\tau)+r \,
    \left(u^\alpha(\tau)+w^\alpha \left( \tau, \theta, \phi \right) \right) \, .
\end{equation}
Next, the new coordinates are used to parametrize the Liénard-Wiechert
potential $A^\alpha$, the field strength tensor
$F^{\alpha\beta}=\partial^\alpha A^\beta-\partial^\beta A^\alpha$, and
the energy-momentum tensor $4 \pi T^{\alpha\beta}=
F^{\alpha\gamma}F_\gamma^\beta+\frac{1}{4}\eta^{\alpha\beta}
F^{\gamma\delta} F_{\gamma\delta}$. Furthermore, the parameterizations of
the tube $V(\tau_1,\tau_2)$ and its normal vector $n_\beta(x)$ are
introduced . 

From here on we will use the light-cone coordinates without further
notice. For the sake of readability, we will suppress arguments of
functions whenever there is no ambiguity. It is understood that fields
are evaluated at $x^\alpha$, partial derivatives $\partial^\alpha$ are
meant w.r.t.\ argument $x^\alpha$, and four-vectors derived from the
world-line $z^\alpha$ of the charge are evaluated at $\tau$. 

\subsection{The energy-momentum tensor}
 The Liénard-Wiechert potential is given by
\begin{equation}
A^\alpha=q\frac{u^\alpha}{r} \, .
\end{equation}
In several occasions, for example for the field strength
tensor, the derivatives
\begin{equation}\label{fete}
\partial^{\alpha}A^{\beta}=q\left(\frac{a^{\beta}\partial^{\alpha}\tau}{r}-\frac{u^{\beta}\partial^{\alpha}r}{r^2}\right)
\end{equation}
 need to be calculated. Hence,
\begin{eqnarray}\label{pr}
&&\partial^{\alpha}r=\partial^{\alpha}((x^\beta-z^\beta)u_\beta)\nonumber\\
&&=u^\alpha+x^\beta a_\beta\partial^{\alpha}\tau-u^\beta u_\beta \partial^{\alpha}\tau-z^\beta a_\beta\partial^{\alpha}\tau
\end{eqnarray}
and $\partial^{\beta}\tau$  are needed. The defining relation of the retarded time
$(x^\alpha-z^\alpha)(x_\alpha-z_\alpha)=0$
 is employed to compute 
\begin{eqnarray}
&&\partial^{\beta}(x^\alpha x_\alpha-2x^\alpha z_\alpha+z^\alpha z_\alpha)=0 \, , \\
&&2x^\beta-2z^\beta-2x^\alpha u_\alpha \partial^{\beta}\tau+2z^\alpha u_\alpha \partial^{\beta}\tau=0 \, , \\
\label{ptau}
&&\partial^{\beta}\tau=\frac{x^\beta-z^\beta}{(x^\alpha-z^\alpha)u_\alpha}=\frac{x^\beta-z^\beta}{r}=u^\beta+w^\beta \, .
\end{eqnarray}
For the field strength tensor the abbreviation $a^\alpha_\perp=a^\alpha+a^\beta
w_\beta w^\alpha$ is used, which is orthogonal to the vectors $u^\alpha,
w^\alpha$ implying $a^\alpha_\perp w_\alpha=0$ and $a^\alpha_\perp u_\alpha=0$. The
field strength tensor $F^{\alpha\beta}=\partial^\alpha A^\beta-\partial^\beta
A^\alpha $ is then given by
\begin{eqnarray}
\label{field-strength-tensor}
F^{\alpha\beta}=&&\frac{q}{r^2}(w^\alpha u^\beta-u^\alpha w^\beta)\nonumber\\
&&+\frac{q}{r}((u^\alpha+w^\alpha)a^\beta_\perp-a^\alpha_\perp(u^\beta+w^\beta)) \, .
\end{eqnarray}
The first line in (\ref{field-strength-tensor}) is the boosted
Coulomb field contribution and the
second the radiation field contribution. The associated energy-momentum tensor $4 \pi
T^{\alpha\beta}=F^{\alpha\gamma}F_\gamma^\beta+\frac{1}{4}\eta^{\alpha\beta}
F^{\gamma\delta} F_{\gamma\delta}$ is given by
\begin{eqnarray}\label{emten}
4 \pi T^{\alpha\beta}&&=\frac{q^2}{r^4}(u^{\alpha}u^{\beta}-w^{\alpha}w^{\beta}-\frac{1}{2}\eta^{\alpha\beta}) \nonumber\\
&& +\frac{q^2}{r^3}(a_\perp ^{\beta}(u^{\alpha}+ w^{\alpha})+a_\perp ^{\alpha}(u^{\beta}+ w^{\beta}))\nonumber\\
&&-\frac{q^2}{r^2}a_\perp ^{\gamma}a_{\perp \gamma}(u^{\alpha}+ w^{\alpha})(u^{\beta}+ w^{\beta}) \, .
\end{eqnarray}
The derivation of the expressions (\ref{lightcone}),
(\ref{field-strength-tensor}), (\ref{emten}) and the coordinates can
be found in \cite{parrott2012relativistic} or \cite{rohrlich2007classical}.

\subsection{Parametrization of the tube 
$V(\tau_1,\tau_2)$  in light cone coordinates}
As an introduction we first review how Parrott and Dirac define their
tubes. Both use an implicit definition over the lateral surface of
their tubes. 

The explicit expressions for those 3-dimensional lateral hyper-surfaces
is given in terms of the coordinates $\tau,\theta$, and
$\phi$, while $\tau_1 \leq \tau \leq \tau_2$. Setting $r=\epsilon$ the
lateral surface of Parrott's tube is obtained
\begin{equation}
t^\alpha(\tau,\theta,\phi)=z^\alpha(\tau)+\epsilon(u^\alpha(\tau)
+w^\alpha(\tau,\theta,\phi))\, ,
\end{equation}
which is one of the simplest and at first sight natural
choices first employed by Bhabha. The big advantage is
that in the limit $\tau_1\rightarrow\tau_2$ the retarded time $\tau$
for all points on this surface is the same. The disadvantage is that
an integration over this area does not lead to a total force, as has
been discussed in chapter \ref{capform}.
\begin{figure}
	\includegraphics[height=5cm]{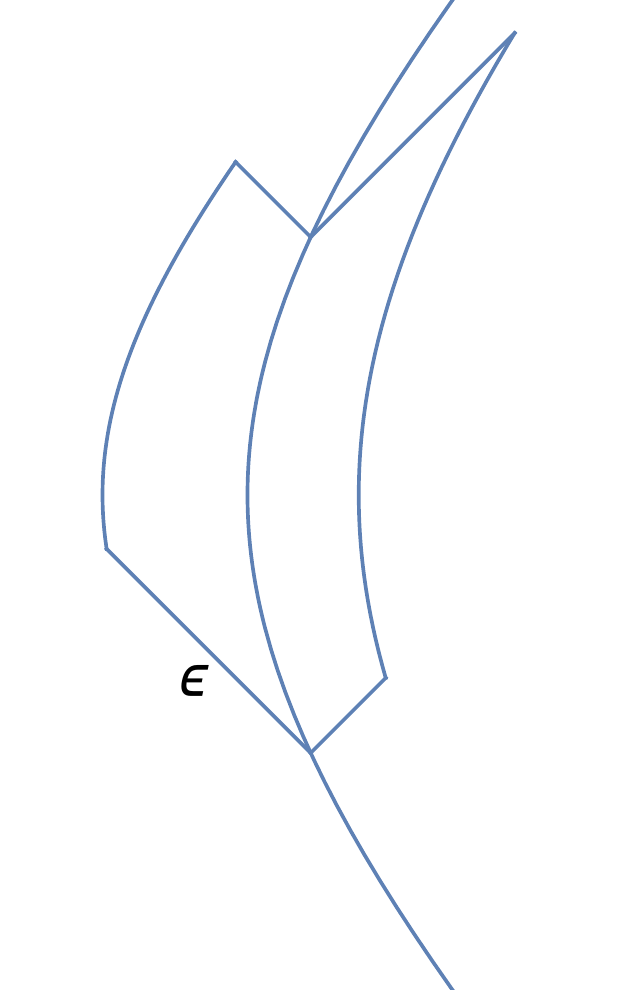}
	\caption{\label{tubeparrott} Parrott's tube, it is defined by
          moving the length $\epsilon$ along the light cone in all
          directions between $\tau_1$ and $\tau_2$.}
\end{figure}
The lateral surface of Dirac's tube is given by
\begin{equation}
t^\alpha(\tau,\theta,\phi)=z^\alpha(\tau)+\epsilon w^\alpha(\tau,\theta,\phi)\, .
\end{equation}
One has to be careful with the meaning of the argument $\tau$ here,
since the retarded time corresponding to some point $t^\alpha$ is not
$\tau$. This is the case because this representation of the surface
does not respect the usual form of light cone coordinates
(\ref{lightcone}). The advantage is, however, that the caps are
hyperplanes of simultaneity in the rest frames as necessary for the
integration.

The choice of $V(\tau_1,\tau_2)$ used in our derivation originates from
the ones of Dirac and Parrott.
We now parametrize Dirac's cap at $\tau_1$ in such a way that $\tau$ still is
the retarded time. An arbitrary point $x^\alpha$ lies in the cap if and only if
the vector $x^\alpha-z^\alpha(\tau_1)$ is orthogonal to the normal vector of
the cap. The normal vector is nothing else than $u^\alpha(\tau_1)$. So we
demand
\begin{equation}
(x^\alpha-z^\alpha(\tau_1)) \, u_\alpha(\tau_1)=0 \, .
\end{equation}
Next we use the light cone coordinates (\ref{lightcone}) for
$x^\alpha$. We follow Parrott's approach and treat the radius $r$ not
as a coordinate but as some function of the coordinates $\tau$,
$\theta$, and $\phi$. This leads to the equation 
\begin{equation}
(z^\alpha+r(u^\alpha+w^\alpha)-z^\alpha(\tau_1))u_\alpha(\tau_1)=0
\end{equation} 
for $r$. The result is
\begin{equation}\label{rad}
r=\frac{(z^\alpha(\tau_1)-z^\alpha)u_\alpha(\tau_1)}{(u^\alpha+w^\alpha)u_\alpha(\tau_1)} \, .
\end{equation}
We now have the desired parametrization for the cap
\begin{equation}
\label{cap}
c^\alpha(\tau,\theta,\phi)=z^\alpha+\frac{(z^\alpha(\tau_1)-z^\alpha)u_\alpha(\tau_1)}
{(u^\alpha+w^\alpha)u_\alpha(\tau_1)}(u^\alpha+w^\alpha) \, ,
\end{equation}
where $\tau_1-\epsilon \leq \tau \leq \tau_1$. The next step is to
find a tube which has such caps. The easiest way is to connect two
caps by a smooth transformation. 
The boundary of the cap at $\tau_1$ is defined by $\tau=\tau_1
-\epsilon$. By shifting $\tau_1$ to $\tau_2$ the desired hyper-surface
is obtained. All that has to be done is to replace $\tau_1-\epsilon$ by $\tau$ in
(\ref{cap}). With this replacement we get the equation for the tube
surface
\begin{eqnarray}\label{tube}
&&t^\alpha(\tau,\theta,\phi)=\nonumber\\
&&z^\alpha+\frac{(z^\alpha(\tau+\epsilon)-z^\alpha)u_\alpha(\tau+\epsilon)}{(u^\alpha+w^\alpha)u_\alpha(\tau+\epsilon)}(u^\alpha+w^\alpha)
   \, ,
\end{eqnarray}
where now $\tau_1-\epsilon \leq \tau \leq \tau_2-\epsilon$ holds.
\begin{figure}
	\includegraphics[height=5cm]{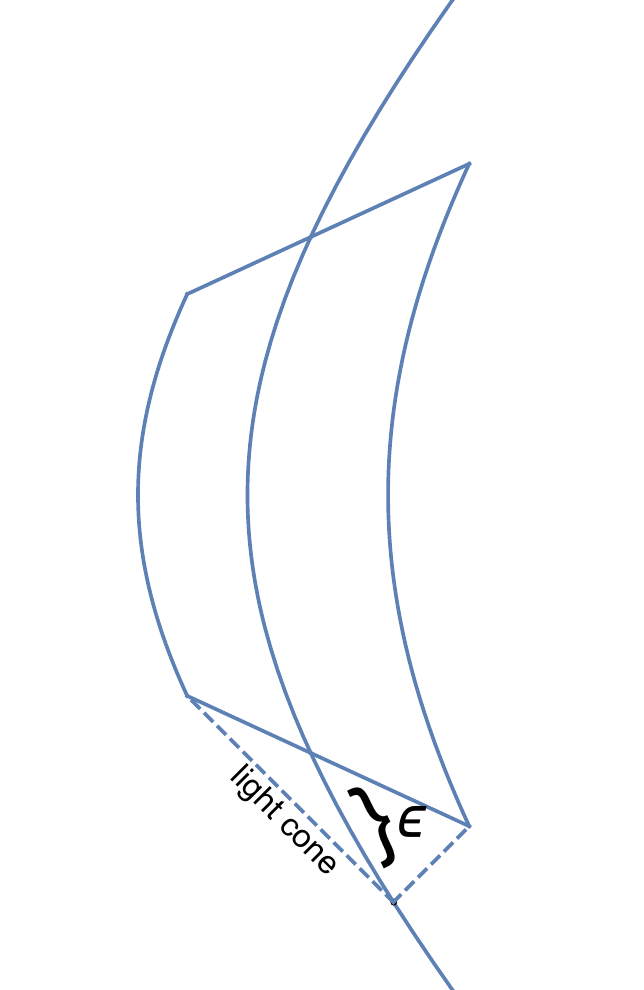}
	\caption{\label{tubenew} Our tube, it is defined first taking
          the cut between the hyperplane of simultaneity in the rest
          frame at $z^\alpha(\tau)$ and the forward light cone
          originating from $z^\alpha(\tau-\epsilon)$ and second taking
          the union of all those cuts between $\tau_1$ and $\tau_2$.}
\end{figure} 
As a word of caution it has to be mentioned that the definition of the
tube surface breaks down for to high accelerations.
Hyperplanes of simultaneity in the rest frame at different times
always intersect somewhere if the velocities are different at those
times. It can happen that this intersection area is closer to the
world-line than the radius $\epsilon$ if the acceleration is bigger than
$1/\epsilon$ between these times. This is a general phenomenon in
special relativity and not specific to our definitions.

The actual calculation of the energy-momentum flow through the tube
(\ref{tube}) is rather long, so it is given in the appendix
\ref{appendix}. The flow is obtained by first evaluating
an explicit expression for the normal vector on the tube (\ref{tube})
and second integrating the contraction of the energy-momentum tensor
(\ref{emten}) and the normal vector over tube surface as in
(\ref{gaus}). The next chapter discusses the result of this
calculation.

\subsection{The new radiation reaction force}
The radiation reaction force is given by 
\begin{eqnarray}\label{endgl}
\partial_\tau P^\alpha_\epsilon (\tau)
= &&-\frac{q^2}{6 \left[ \left( z^\gamma-z^\gamma
	(\tau-\epsilon) \right) \, u_\gamma \right]^2}\nonumber\\
&&\left\{ \left[ u^\alpha-4u^\alpha(\tau-\epsilon) \, u^\beta
u_\beta(\tau-\epsilon) \right] \right.\nonumber \\
&&\left. \times \left[ 1 - u^\delta u_\delta(\tau-\epsilon)+\left( z^\rho-z^\rho
(\tau-\epsilon) \right) \, a_\rho \right] \right\} \nonumber\\
&&-
\frac{q^2}{6 \left( z^\gamma-z^\gamma(\tau-\epsilon) \right) \,
	u_\gamma} \nonumber \\
&&\left\{ 4u^\alpha
(\tau-\epsilon) \left[ a^\tau(\tau-\epsilon) \, u_\tau+u^\zeta
(\tau-\epsilon) \, a_\zeta \right] \right. \nonumber \\
&&\left. +4a^\alpha(\tau-\epsilon) \, u^\vartheta(\tau-\epsilon) \,
u_\vartheta-a^\alpha \right\} \nonumber\\
&& +\frac{2q^2}{3} \, a^\varphi(\tau-\epsilon) \, a_\varphi(\tau-\epsilon) \,
u^\alpha(\tau-\epsilon) \\
=: L^\alpha_\epsilon(\tau) &&\, .
\end{eqnarray}

To our knowledge, $L^\alpha_\epsilon(\tau)$ is the first explicit expression for the
radiation reaction force for an extended charged particle in contrast to the
approximations in terms of Taylor series in $\epsilon$ which, as we have
argued, can be a source dynamical instability.

Nevertheless, it is interesting to perform a Taylor expansion nonetheless in
order to see if our expression, at least in lowest orders of $\epsilon$, agrees
with the right-hand side of the LAD equation which as found in various
computations in the classical literature.

To do this, we make use of $u^\alpha u_\alpha=1$,  $a^\alpha
u_\alpha=0$, and $\dot{a}^\alpha u_\alpha=-a^\alpha a_\alpha$. For the terms
starting with the first fraction in (\ref{endgl}), it is enough to examine the
following bracket
\begin{equation}
1-u^\delta u_\delta(\tau-\epsilon)+(z^\rho-z^\rho
(\tau-\epsilon)) \, a_\rho =\mathcal{O}(\epsilon^3) \, .
\end{equation}
This fraction does not contribute in the limit $\epsilon\to 0$ term since the
nominator is of order $\mathcal O (\epsilon^2)$. The first bracket following
the second fraction also does not contribute in this limit since
\begin{equation}
4u^\alpha (\tau-\epsilon) \left[ a^\tau(\tau-\epsilon) \, u_\tau+u^\zeta
(\tau-\epsilon) \, a_\zeta \right]=\mathcal{O}(\epsilon^2) \, .
\end{equation}
The remaining terms reduce to the well-known LAD force
\begin{eqnarray}
    (\ref{endgl})=-&&\frac{q^2}{6}\Big(\frac{4a^\alpha(\tau-\epsilon)
   u^\vartheta(\tau-\epsilon) \,
   u_\vartheta-a^\alpha}{\left( z^\gamma-z^\gamma(\tau-\epsilon) \right) \, u_\gamma}
   \nonumber\\
                  && \phantom{\frac{q^2}{6}\Big(}-4 a^\varphi(\tau-\epsilon)a_\varphi(\tau-\epsilon)
                 u^\alpha(\tau-\epsilon)\Big) \nonumber\\
=&&-\frac{q^2}{2\epsilon}a^\alpha+\frac{2q^2}{3}(\dot{a}^\alpha+a^\varphi
                                                            a_\varphi
                                                            u^\alpha)+\mathcal{O}(\epsilon)
                                                            \, ,
                                                            \label{LADeq}
\end{eqnarray}
which requires a mass renormalization procedure to get rid of the first
$O(\epsilon^{-1})$ term by absorbing it into the inertial mass.  However, as
argued above, this expansion in $\epsilon$ is not helpful for arriving at a
sensible dynamics as it is the source of dynamical instabilities.

\section{\label{eff-eqn}Effective equations of motions}
In this final chapter we draw from the previously establish result
\eqref{endgl} which describes the change of total momentum of our charge
distribution. In order to formulate a self-consistent dynamics we still need to
establish a relation between this change of momentum and the corresponding
change of velocity of the world-line $\tau\mapsto z^\alpha(\tau)$. Here, we
face the problem that $P^\alpha_\epsilon(\tau)$ is the total change of momentum
of the charge distribution defined by our $\epsilon$-depending tube, as given in
Figure~\ref{tubenew}. In fact, at this point we would need to compute
$j_\epsilon^\alpha(x)$ from \eqref{j} and establish the desired relation in view of
Dirac's argument \eqref{gaus} by
\begin{align}
    \label{momentum-change}
    L^\alpha_\epsilon(\tau) 
    =\partial_\tau P_\epsilon^\alpha(\tau)=\partial_\tau \int_{V(\tau_1,\tau)}
    d^4x\, F^{\alpha\beta}_\epsilon(x)
    {j_{\epsilon}}_\beta(x).
\end{align}
Note that for point-charges the right-hand side simply reduces to the
well-known Lorentz force exerted by the electromagnetic field on the charge;
see \eqref{gaus}.  In order to avoid this step we make the following model
assumption:
\begin{align}
    \label{model-assumption}
    P^\alpha_\epsilon(\tau) = m_\epsilon(\tau) \dot z^\alpha(\tau)
\end{align}
for $m_\epsilon(\tau)$ being a proportionality factor that for sake of
generality may depend on $\tau$. At this point the explicit dependence on
$\tau$ may appear strange, however, it will turn out that this additional
degree of freedom will be helpful to arrive at a concept of total inertial mass
taking account of the one that is effectively created by the back reaction
of the electromagnetic field.  By contracting  the equality 
\begin{align}
    L_\epsilon^\alpha(\tau)
    = 
    \partial_t m_\epsilon(\tau) \dot z^\alpha(\tau)
    +
    m_\epsilon(\tau) \ddot z^\alpha(\tau)
\end{align}
with $\dot z_\alpha(\tau)$ and, exploiting $\dot z_\alpha(\tau) \, \dot z^\alpha(\tau)=1$ and $\dot
z_\alpha(\tau) \, \ddot z^\alpha(\tau)=0$, we infer
\begin{align}
    \partial_t m_\epsilon(\tau) = L^\alpha_\epsilon(\tau)\dot z_\alpha(\tau)\, .
\end{align}
Defining the relativistic force 
\begin{align}
    F_\epsilon^\alpha(\tau)
    :=  L^\alpha_\epsilon(\tau) - \dot z^\alpha L_\epsilon^\beta(\tau)\dot z_\beta(\tau)
\end{align}
that is four-orthogonal to $\dot z^\alpha(\tau)$, we arrive at the dynamical
system
\begin{align}\label{forceeqn}
    \frac{d}{d\tau}
    \begin{pmatrix}
        z^\alpha_\epsilon (\tau) \\
        u^\alpha_\epsilon (\tau) \\
        m_\epsilon(\tau)
    \end{pmatrix}
    =
    \begin{pmatrix}
        u^\alpha_\epsilon (\tau) \\
       \frac{1}{ m_\epsilon(\tau)} \, \left( F^\alpha_\epsilon[z](\tau)
            +
        F^\alpha_\text{ext}(\tau) \right) \\
        {u_\epsilon}_\alpha(\tau) L^\alpha_\epsilon[z](\tau)
    \end{pmatrix} \,, 
\end{align}
where for the discussion below we introduced an additional external force
$F^\alpha_\text{ext}(\tau)$ acting on the charge that is four-orthogonal to
$\dot z^\alpha(\tau)$. This system couples the world-line $\tau\mapsto
z^\alpha(\tau)$ to the change of momentum computed in \eqref{endgl} caused by
the electromagnetic field that, in turn, is produced by the charge itself.
Here, the argument $[z]$ in square brackets is to remind us that these terms are functionals of
the world-line $t\mapsto z^\alpha(\tau)$.  In fact, inspecting the expression
\eqref{endgl} reveals that the system \eqref{forceeqn} effectively turns out to
be a system non-linear and neutral delay equations. The delay stems form the
fact that the charge has the extension of our $\epsilon$ tube and the speed of
light is finite and is therefore expected. Note that the initial value of the
proportionality factor $m(0)$ is an additional degree of freedom. Based on the
general theory of delay equations, it is to be expected that the initial values
of this system are twice continuously differentiable
trajectory strips $z^\alpha:[-\epsilon,0]\to\mathbb R^4$ together with the
value $m_\epsilon(0)\in\mathbb R$.

To understand this system of equations \eqref{forceeqn} better, we consider the
simple case of an external force $F^\alpha_\text{ext}$ that is tuned to force
the charge into an uniform acceleration, say, along the $z$-coordinate for
$\tau \in\Lambda$. If it was not for the expected radiation reaction force we
are interested in, this setting could be thought of a charge in a constant
electric field. Here, however, it is important to keep in mind that the
considered external force also compensates possible friction, e.g., due to the
radiation reaction, to keep the acceleration constant. In this case the world-line is
given by
\begin{eqnarray}
    z^\alpha(\tau)=\frac{1}{g} \, \left( \sinh (g \tau), 0, 0, \cosh (g\tau) \right) \, ,
\end{eqnarray}
where $g$ is the constant acceleration on the interval $\tau\in\Lambda$. The 
change in momentum due to the back reaction, computed from (\ref{endgl}), has
the correspondingly simple form
\begin{equation}\label{uniforce-new}
L_\epsilon^\alpha(\tau) = -\frac{q^2g}{2\sinh(g\epsilon)} \, a^{\alpha}(\tau)
\end{equation}
for $\tau\in\Lambda$. In view of the equation of motion \eqref{forceeqn} we
obtain $\dot m_\epsilon(\tau)=0$, and hence,
\begin{eqnarray}
\label{time-dependent-mass}
&&\left( m_\epsilon(0) + \frac{q^2g}{2\sinh(g\epsilon)} \right)
   a^{\alpha} (\tau) = 
F^\alpha_\text{ext}(\tau)\, ,
\end{eqnarray}
which gives rise to the following total inertial mass when measured w.r.t.\ the
external force:
\begin{eqnarray}
\label{meff}
m_{\text{tot}} = m_\epsilon(0) + \frac{q^2g}{2\sinh(g \epsilon)} \, .
\end{eqnarray}
Two properties of (\ref{meff}) can be observed: First, as in \eqref{LADeq}, the
correction to the inertia originating from the electromagnetic field in leading
order as $\epsilon\to 0$ equals $q^2/2\epsilon$:
\begin{eqnarray}
    \label{mass_tot}
    m_{\text{tot}} = m_\epsilon(0) + \frac{q^2}{2\epsilon} -
    \frac{q^2g^2}{12}  \, \epsilon + O(\epsilon^2) \, . 
\end{eqnarray}

And second, the higher-order corrections in (\ref{mass_tot}) explicitly depend
on the dynamics itself, in this case on the acceleration parameter $g$. To
illustrate the dependence on the dynamics even more clearly, we consider
eigentimes $0<\tau_1<\tau_2$ with $\tau_2-\tau_1>2\epsilon$ and define
$\Lambda_1=[0,\tau_1)$, $\Lambda=[\tau_1,\tau_2)$, and
$\Lambda_2=[\tau_2,+\infty)$.  We assume that the external force
$F^\alpha_\text{ext}$ is tuned such that the effective acceleration in
$\Lambda_1\cup[\tau_1,\tau_1+\epsilon)$ and $[\tau_2-\epsilon)\cup\Lambda_2$ is
constant and equal to $g_1$ and $g_2$, respectively, where $g_1\neq g_2$.
Furthermore, we assume that in the intermediate interval $[\tau_1+\epsilon,
\tau_2-\epsilon)$ the acceleration of the charge changes smoothly from $g_1$ to
$g_2$ obeying $\dot{m}_\epsilon (\tau) = 0$ for $\tau\in\Lambda$. By virtue of
\eqref{forceeqn} we observe that
\begin{align}
    z_i^\alpha(\tau)&=\frac{1}{g_i} \, \left( \sinh (g_i \tau), 0, 0, \cosh (g_i\tau)
    \right) \quad \text{for }\tau\in\Lambda_i
\end{align}
solves \eqref{forceeqn} for $i=1,2$ which, by \eqref{meff}, implies that the
corresponding total inertial mass $m_\text{tot}$ depends on the eigentime
$\tau$, more precisely it holds
\begin{align}
    m_\text{tot} =
    m_\epsilon(0)+
    \begin{cases}
        \frac{q^2g_1}{2\sinh(g_1\epsilon)} \quad \text{for } \tau\in\Lambda_1\\
        \frac{q^2g_2}{2\sinh(g_2\epsilon)} \quad \text{for } \tau\in\Lambda_2
    \end{cases} \, .
\end{align}
It is therefore to be expected that the total inertial mass is a dynamical
quantity. The concept of a time-dependent total inertial mass is not new but
also observed in other theories treating back-reaction, e.g., in
Bopp-Podolski's generalized electrodynamics \cite{zayats2014self}. For our
system \eqref{forceeqn} the time-dependency is foremost due to the
time-dependent shape of our $\epsilon$ tube.\\

\section{Conclusion}

Whether $\epsilon$ is kept finite or a limit $\epsilon\to 0$ is considered, in
our approach the inertial mass is an emerging phenomenon that originates from
the back reaction on the charge exerted by its own electromagnetic field.
Thus, a general procedure is needed to gauge the inertial mass to the one
observed in the experiment. In view of \eqref{mass_tot}, the renormalization
procedure $m_\epsilon(0)=m_\text{exp}-q^2/2\epsilon$ for $m_\text{exp}$ being
the experimentally measured inertial mass, as also employed by Dirac, is
appropriate as long as the time-dependent terms are subleading. However, we
emphasize again that the higher-order terms in \eqref{mass_tot} may not simply
be neglected in a limiting procedure $\epsilon\to 0$ as the Taylor expansion of
solutions $z^\alpha(\tau)$ on the right-hand side of the equations of motion
\eqref{forceeqn} cannot be controlled uniformly on time intervals. 
The neglect of higher-order terms may provoke the so-called runaway
solutions as illustrated with the counter example given in
Section~\ref{taylor}. The virtue of our approach is therefore that no Taylor
expansion has been employed when formulating the law of motion
\eqref{forceeqn}. Instead we are left with an explicit expression
\eqref{endgl} that can readily be studied analytically or numerically
in various settings. One imminent question is whether the dynamical
system \eqref{forceeqn} is stable and does in particular not lead to
the notorious run-away solutions. A thorough analysis of this question
is left for a forthcoming paper.

The only assumptions
involved in the derivation of system \eqref{forceeqn} were:
\begin{enumerate}
    \item Energy-momentum conservation between the kinetic and the field
        degrees of freedom as expressed in differential form in change of
        momentum as given by \eqref{momentum-change}.
    \item The special form of the $\epsilon$-tube that allowes the explicit
        evaluation of the integrals involved in computing the momentum change
        in Section~\ref{final-result}.
    \item The assumption \eqref{model-assumption} that allows to relate
        the change of momentum to the change of velocity which is a pathology
        of the extended charge model.
\end{enumerate}
While assumption 1.\ seems rather natural, assumption 2.\ arises out of the
mathematical necessity to introduce a cut-off in the electromagnetic fields as
the solutions of the Maxwell equations are ill-defined on the world-line for
point-charges.  Of course, in other settings, as the above mentioned
generalized electrodynamics, this point can potentially be avoided at the
cost of replacing Maxwell's equations with a more regular version of
the latter. This may be a valid approach but is not our focus
here. Moreover, one may wonder, how much information of the particular
shape of the employed $\epsilon$-tube enters the law of motion
\eqref{forceeqn}. In view of the Stoke's theorem employed in
the derivation of the momentum change, recall \eqref{gaus}, only the geometric
properties of the caps of the tube enter in expression \eqref{endgl}.
Assumption 3.\ is certainly the most ad hoc one.  Indeed, a more subtle
analysis of \eqref{momentum-change} is required to argue for the validity of
the given approximation \eqref{model-assumption} in a certain regime.
However, this goes beyond the scope of this work. Furthermore, the
explicit form of the law of motion \eqref{forceeqn} allows the
exploration of example settings, such as the synchrotron setting, in
which a charge moves in a constant magnetic field perpendicular to the
motion, for which already other approaches, such as the
Landau-Lifschitz equations, make predictions. Based on an
understanding in these settings, a sensible renormalisation procedure
has to be developed. It is our hope that the additional degree of
freedom in $m_\epsilon(\tau)$ can compensate the time-dependencies of
our $\epsilon$-tube to some extend so that in a regime of sufficiently
small $\epsilon$ the renormalised solutions to \eqref{forceeqn} become
rather independent of the cut-off. Both of these open points will be addressed
in a follow-up article which is in preparation.

\begin{acknowledgments}
C.B.\ and H.R.\ acknowledge the hospitality of the Arnold Sommerfeld Center at
the Ludwig-Maximilians-Universit\"at in Munich. H.R. is grateful for
discussions with Peter Mulser. C.B. and D.-A.D. are grateful for
fruitful discussions with Stephen Lyles and Michael Kiessling. This
work has been funded by the Deutsche Forschungsgemeinschaft (DFG)
under Grant No. 416699545 within the Research Unit FOR2783/1, under
under Grant No. Ru633/3-1 within the Research Unit TRR18, the cluster
of excellence EXC158 (MAP) and furthermore by the junior research
group ``Interaction between Light and Matter'' of the Elite Network Bavaria.
\end{acknowledgments}

\section{Appendix: Computation of the force}\label{appendix}
\subsection{The normal vector on the tube}
The direct way to calculate the normal vector $n^\alpha$ is to make use of
the fact that there exists only one unit vector which is orthogonal to all
the tangent vectors of the tube $t^\alpha$ up to the sign. The three
tangent vectors are given by the derivatives of $t^\alpha$ with
respect to $\tau$, $\theta$, and $\phi$. The contraction of those
three vectors with the epsilon tensor gives a vector which is
automatically orthogonal to the tangent vectors. In some sense the
epsilon tensor is the generalization of the cross product to higher
dimensions. In the language of differential geometry, the normal vector
$n^\alpha$ is the hodge dual of the wedge product of the tangent
vectors. It follows that the normal vector $n^\alpha$ is given by
\begin{equation}\label{normalvec}
n^\alpha=\epsilon^{\alpha\beta\gamma\delta} \, \partial_\tau t_\beta
\, \partial_\theta t_\gamma  \, \partial_\phi t_\delta \, .
\end{equation}
Its length is the volume spanned by a unit normal vector and the three
tangent vectors, which is nothing else than the Jacobian
determinant. Hence, we do not even need to adjust it. The tangent vectors
are
\begin{eqnarray}
&&\partial_\tau t^\alpha=u^\alpha+\partial_\tau r(u^\alpha+w^\alpha)
+r(a^\alpha+\partial_\tau w^\alpha) \, , \\
&&\partial_\theta t^\alpha=\partial_\theta r(u^\alpha+w^\alpha)
+r(\partial_\theta w^\alpha) \, , \\
&&\partial_\phi t^\alpha=\partial_\phi r(u^\alpha+w^\alpha)
+r(\partial_\phi w^\alpha) \, . 
\end{eqnarray}
The derivatives of r are lengthy expressions. So it makes sense not to
calculate a complete expression for the normal vector but instead
state only its components in an useful orthonormal basis. For this
basis we choose 
\begin{eqnarray}
&&u^\alpha \, , \quad w^\alpha \, , \quad
\theta^\alpha=\partial_\theta w^\alpha \, , \quad
\phi^\alpha=\frac{\partial_\phi w^\alpha}{\sin \theta}
\end{eqnarray}
and
\begin{eqnarray}
\label{basis-expansion}
&&n^\alpha=n^\beta u_\beta u^\alpha-n^\beta w_\beta w^\alpha \nonumber\\
&&\hspace{2cm}-n^\beta\theta_\beta \theta^\alpha-n^\beta \phi_\beta \phi^\alpha \, .
\end{eqnarray}
We start with the term $n^\beta u_\beta$ on the right-hand side of
(\ref{basis-expansion}) which yields\\
\begin{eqnarray}\label{nu}
n_\delta u^\delta&=&\epsilon_{\alpha\beta\gamma\delta} \partial_\tau
t^\alpha \partial_\theta t^\beta \partial_\phi t^\gamma
u^\delta \nonumber\\
\label{u-n}
&=&\begin{vmatrix}
1+\partial_\tau r -r a^\beta w_\beta & \partial_\theta
r& \partial_\phi r&1\\\partial_\tau r-r a^\beta w_\beta
& \partial_\theta r& \partial_\phi r&0\\ -r a^\beta
\theta_\beta-r \partial_\tau  w^\beta \theta_\beta & r &0&0\\-r
a^\beta \phi_\beta-r \partial_\tau  w^\beta \phi_\beta&0&\sin \theta
\,r&0
\end{vmatrix}\nonumber\\
&=& -(r^2 \sin \theta \, (\partial_\tau r -r a^\beta w_\beta) \nonumber \\
&&+r^2\partial_\phi r(a^\beta \phi_\beta+ \partial_\tau  w^\beta \phi_\beta)\nonumber\\
&&+r^2\sin \theta \,\partial_\theta r(a^\beta
\theta_\beta+ \partial_\tau  w^\beta \theta_\beta)) \, ,
\end{eqnarray}
where use has been made of $u^\alpha w_\alpha=0$ leading to
$u^\alpha \partial_\tau w_\alpha=-a^\alpha w_\alpha$. To obtain the
term $n^\beta w_\beta $ in (\ref{basis-expansion}) the $1$ showing up
in the last column of the determinant in (\ref{u-n}) has to shifted
down by one row. This yields
\begin{eqnarray}
n_\alpha w^\alpha&=&r^2 \sin \theta \, (1+\partial_\tau r-r a^\beta w_\beta)\nonumber\\
&&+r^2\partial_\phi r(a^\beta \phi_\beta+ \partial_\tau  w^\beta \phi_\beta)\nonumber\\
&&+r^2\sin \theta \,\partial_\theta r(a^\beta \theta_\beta+ \partial_\tau  w^\beta \theta_\beta) \, .
\end{eqnarray}
The contractions of $n_\alpha$ with $\theta^\alpha$ and $\phi^\alpha$
are obtained in the same way by shifting the $1$ further down. This
gives
\begin{eqnarray}
&&n_\alpha  \theta^\alpha=-r \sin \theta \, \partial_\theta r \, , \\
&&n_\alpha  \phi^\alpha=-r  \partial_\phi r \, .\label{nphi}
\end{eqnarray}
The derivatives of the radius in (\ref{tube}) are given by
\begin{eqnarray}
\partial_\tau r&=&\Big(\big[ (u^\alpha(\tau+\epsilon)-u^\alpha) \, u_\alpha(\tau+\epsilon)
\nonumber\\
&&\hspace{1cm}+(z^\alpha(\tau+\epsilon)-z^\alpha) \, a_\alpha(\tau+\epsilon) \big]
\nonumber \\
&&\hspace{3cm}\times \left( u^\beta+w^\beta \right) \, u_\beta(\tau+\epsilon ) \nonumber \\
&&-\left( z^\beta(\tau+\epsilon)-z^\beta \right) \,
u_\beta(\tau+\epsilon) \nonumber \\
&&\hspace{1cm}\times \big[ \left( a^\alpha+\partial_\tau w^\alpha
\right) \, u_\alpha(\tau+\epsilon) \nonumber\\
&&\hspace{2cm}+\left( u^\alpha+w^\alpha \right) \, a_\alpha(\tau+\epsilon) \big] \Big) \nonumber \\
&&/\big[ \left( u^\alpha+w^\alpha \right) \, u_\alpha(\tau+\epsilon) \big]^2 \, , \\
\partial_\theta r&=&-\left( z^\alpha(\tau+\epsilon)-z^\alpha \right)
\, u_\alpha(\tau+\epsilon) \nonumber \\
&&\hspace{1cm}\times \left[ \partial_\theta w^\beta \, 
u_\beta(\tau+\epsilon) \right] \nonumber \\
&&/\big[ \left( u^\alpha+w^\alpha \right) \, u_\alpha(\tau+\epsilon) \big]^2 \, ,
\\
\partial_\phi r&=&-\left( z^\alpha(\tau+\epsilon)-z^\alpha \right)
\, u_\alpha(\tau+\epsilon) \nonumber \\
&&\hspace{1cm}\times \left[ \partial_\phi w^\beta \, 
u_\beta(\tau+\epsilon) \right] \nonumber \\
&&/\big[ (u^\alpha+w^\alpha) \, u_\alpha(\tau+\epsilon) \big]^2 \, .
\end{eqnarray}
Now the contraction of the energy-momentum tensor (\ref{emten}) with
the normal vector $n^\alpha$ can be calculated. The corresponding
calculations and integrations are carried out in the next section.

\subsection{Computation of the change of the momentum}\label{final-result}
We start with (\ref{gaus}) and the domain of (\ref{tube}) to obtain
\begin{eqnarray}\label{totalint}
\partial_\tau P^\alpha_\epsilon(\tau) &=&- \partial_\tau \int_{\partial
	V(\tau_1,\tau)} d^3x_\beta
T^{\alpha\beta}_\epsilon \nonumber\\
&=&-\partial_\tau
\int_{\tau_1-\epsilon}^{\tau-\epsilon} d\tau \int_0^\pi
d\theta\int_0^{2\pi} d\phi \, n_\beta \,T^{\alpha\beta} \, .
\end{eqnarray}
In the following we also consider the integral domains given in
\eqref{totalint} and suppress their reference in our notation.
Due to the cut-off, the cap integrals vanish since
$T^{\alpha \beta}_\epsilon=0$ within the tube and only the
integral over the lateral surface of the tube remains, where
$T^{\alpha\beta}_\epsilon=T^{\alpha\beta}$ holds. First the angle
integration is performed and also a factor of $4\pi/q^2$ is introduced
for convenience. To carry out the calculations, we make use of
$\eta^{\alpha\beta}=u^\alpha u^\beta-w^\alpha w^\beta-\theta^\alpha
\theta^\beta-\phi^\alpha \phi^\beta$ and (\ref{emten}) for
$T^{\alpha\beta}$ and (\ref{nu})-(\ref{nphi}) for $n_\beta$. This leads to
\begin{widetext}
	\begin{eqnarray}
	\label{long-eqn}
	&&\int d\theta d\phi \, n_\beta \, \frac{4\pi}{q^2}T^{\alpha\beta}\nonumber \\
	&&=\int d\theta d\phi \, n_\beta \, \Big[ \frac{
		u^{\alpha}u^{\beta}-w^{\alpha}w^{\beta}+\theta^\alpha
		\theta^\beta+\phi^\alpha \phi^\beta}{2r^4}+\frac{a_\perp
		^{\beta}(u^{\alpha}+ w^{\alpha})+a_\perp ^{\alpha}(u^{\beta}+
		w^{\beta})}{r^3} -\frac{a_\perp ^{\gamma}a_{\perp \gamma} \left( u^{\alpha}+
		w^{\alpha} \right) \left( u^{\beta}+ w^{\beta}\right)}{r^2}\Big] \nonumber \\
	&&=\int \, d\theta d\phi \, \left( \overbrace{\frac{-u^\alpha \sin \theta
			\,}{2r^2}\partial_\tau r}^i-\overbrace{\frac{w^\alpha \sin \theta
			\,}{2r^2}\partial_\tau r}^{ii}+\overbrace{\frac{u^\alpha \sin
			\theta \,}{2r}a^\beta w_\beta}^{iii}+\overbrace{\frac{w^\alpha \sin
			\theta \,}{2r}a^\beta w_\beta}^{iv}-\overbrace{\frac{w^\alpha \sin
			\theta \,}{2r^2}}^{v}-\overbrace{\frac{u^\alpha}{2r^2} a^\beta
		\phi_\beta \partial_\phi r}^{via} \right. \nonumber\\
	&&\hspace{2cm}-\overbrace{\frac{u^\alpha\sin \theta \,}{2r^2} a^\beta
		\theta_\beta \partial_\theta
		r}^{vib}-\overbrace{\frac{w^\alpha}{2r^2} a^\beta
		\phi_\beta \partial_\phi r}^{viia}-\overbrace{\frac{w^\alpha\sin
			\theta \,}{2r^2} a^\beta \theta_\beta \partial_\theta r}^{viib}
	-\overbrace{\frac{u^\alpha}{2r^2} \partial_\tau w^\beta
		\phi_\beta \partial_\phi r}^{viiia}-\overbrace{\frac{u^\alpha\sin
			\theta \,}{2r^2} \partial_\tau w^\beta \theta_\beta \partial_\theta
		r}^{viiib} \nonumber \\
	&&\hspace{3cm}-\overbrace{\frac{w^\alpha}{2r^2} \partial_\tau w^\beta 
		\phi_\beta \partial_\phi r}^{ixa} -\overbrace{\frac{w^\alpha\sin
			\theta \,}{2r^2} \partial_\tau
		w^\beta \theta_\beta \partial_\theta r}^{ixb}-\overbrace{\frac{\sin
			\theta \, \theta^\alpha}{2r^3}\partial_\theta
		r}^{xa}-\overbrace{\frac{ \phi^\alpha}{2r^3}\partial_\phi
		r}^{xb}+\overbrace{\frac{\sin \theta
			\,a^\alpha_\perp}{r}}^{xi} \nonumber \\
	&&\left. \hspace{5cm}+\overbrace{\frac{\sin \theta 
			\,(u^\alpha+w^\alpha)}{r^2}a^\beta \theta_\beta \partial_\theta 
		r}^{xiia}+\overbrace{\frac{u^\alpha+w^\alpha}{r^2}a^\beta 
		\phi_\beta \partial_\phi r}^{xiib} -\overbrace{a_\perp
		^{\gamma}a_{\perp \gamma}(u^{\alpha}+ w^{\alpha})\sin \theta
	}^{xiii} \right) \, .
	\end{eqnarray}
\end{widetext}
As is seen from (\ref{long-eqn}) only angle integrations remain to be
carried out. Since the integral (\ref{long-eqn}) is a Lorentz vector the
integrations can be carried out in the rest frame. The original
expressions are then obtained by transforming back to lab frame. It is
worth noting that only the vectors $w^\alpha$, $\theta^\alpha$, and
$\phi^\alpha$ depend on the angles $\theta$ and $\phi$.  The quantities
$\theta^\alpha$ and $\phi^\alpha$ only appear in the combination
$\theta^\alpha \theta^\beta+\phi^\alpha \phi^\beta$. In the rest frame 
\begin{equation}
\label{theta-phi}
\theta_0^\alpha \theta_0^\beta+\phi_0^\alpha \phi_0^\beta=
m^{\alpha\beta}-w_0^\alpha w^\beta_0 
\end{equation}
holds, where
\begin{equation}
m^{\alpha\beta}=
\begin{pmatrix}
0&0&0&0\\ 0&1&0&0\\ 0&0&1&0\\ 0&0&0&1\\ 
\end{pmatrix} \, , \quad
w^\alpha_0=
\begin{pmatrix}
0\\ \sin \theta \,\cos \phi \, \\ \sin \theta \,\sin \phi \, \\ \cos \theta 
\end{pmatrix} \, . 
\end{equation}
Making use of (\ref{theta-phi}) and pulling angle independent
terms out of the integrals in (\ref{long-eqn}), all remaining terms are
only integrals over powers of $w^\alpha_0$. The following integrals are needed:
\begin{eqnarray}
\label{rel-int-a}
&&\int d\theta d\phi \, \sin \theta =4\pi \, , \\
&&\int d\theta d\phi \, w_0^\alpha\sin \theta =0 \, , \\
&&\int d\theta d\phi \, w_0^\alpha w_0^\beta\sin \theta =\frac{4\pi}{3} m^{\alpha\beta} \, , \\
&&\int d\theta d\phi \, w_0^\alpha w_0^\beta w_0^\gamma\sin \theta =0 \, , \\
&&\int d\theta d\phi \, w_0^\alpha w_0^\beta w_0^\gamma w_0^\delta\sin \theta \nonumber\\
\label{rel-int-e}
&&=\frac{4\pi}{15} \left( m^{\alpha\beta}m^{\gamma\delta}+m^{\alpha\gamma}m^{\beta\delta}
+m^{\alpha\delta}m^{\gamma\beta} \right) \, . 
\end{eqnarray}
The transformation of $m^{\alpha\beta}$ back to the
lab frame is what remains to be done. To determinate the necessary
Lorentz matrix $\Lambda^\alpha_\beta$ we make use of
\begin{eqnarray}
\label{m-rule}
&&m^{\alpha\beta}=\delta^\alpha_0 \delta^\beta_0-\eta^{\alpha\beta}
\end{eqnarray}
and $\Lambda^\alpha_0=\Lambda^0_\alpha=u^\alpha$. We obtain
\begin{eqnarray}
\label{m-transformed}
&&\Lambda^\alpha_\gamma m^{\gamma\delta}
\Lambda_\delta^\beta \nonumber \\
&&=\Lambda^\alpha_\gamma \, \left( \delta^\gamma_0
\delta^\delta_0-\eta^{\gamma\delta} \right) \, \Lambda_\delta^\beta
\nonumber \\
&&=u^\alpha u^\beta-\eta^{\alpha\beta} \, .
\end{eqnarray}
With the help of (\ref{rel-int-a}) to (\ref{rel-int-e}) and
(\ref{m-transformed}), the integrations $\fbox{i}-\fbox{xiii}$ in (\ref{long-eqn})
are straight forward. We obtain for integral $\fbox{i}$
\begin{eqnarray}
\fbox{i}&=&\int   d\theta d\phi \, \frac{-\sin \theta
	\,u^\alpha}{2r^2} \, \partial_\tau r  \nonumber\\
&=&\int d\theta d\phi \, \frac{-\sin \theta
	\,u^\alpha}{2\left[ (z^\gamma(\tau+\epsilon)-z^\gamma) \,
	u_\gamma(\tau+\epsilon) \right]^2} \nonumber\\
&&\left( \left[ 1-u^\delta
u_\delta(\tau+\epsilon)+\left( z^\mu(\tau+\epsilon)-z^\mu \right) \, a_\mu(\tau+\epsilon)
\right] \right. \nonumber\\
&&\left( u^\beta+w^\beta \right) \, u_\beta(\tau+\epsilon)
-\left( z^\nu(\tau+\epsilon)-z^\nu \right) \,
u_\nu(\tau+\epsilon) \, \nonumber \\
&&\left. \left[ \left( a^\lambda+\partial_\tau
w^\lambda \right) \, u_\lambda(\tau+\epsilon)
+\left( u^\kappa+w^\kappa \right) \, a_\kappa(\tau+\epsilon) \right] \right) \nonumber\\
&=&\frac{-2\pi u^\alpha}{\left[
	\left( z^\gamma(\tau+\epsilon)-z^\gamma \right) \, u_\gamma(\tau+\epsilon) \right]^2}\nonumber\\
&&\left( \left[ 1-u^\delta
u_\delta(\tau+\epsilon)+\left( z^\mu(\tau+\epsilon)-z^\mu \right) \, a_\mu(\tau+\epsilon)
\right] \right. \nonumber\\
&&u^\beta u_\beta(\tau+\epsilon) -\left(
z^\nu(\tau+\epsilon)-z^\nu \right) \, u_\nu(\tau+\epsilon)
\nonumber \\
&&\left. \left[ a^\lambda u_\lambda(\tau+\epsilon) +u^\kappa
a_\kappa(\tau+\epsilon) \right] \right) \, .
\end{eqnarray}
For integral $\fbox{ii}$, we obtain
\begin{eqnarray}
\label{long-int2}
\fbox{ii}&=&\int d\theta d\phi \, \frac{-\sin \theta \,w^\alpha}{2r^2} \, \partial_\tau r  
\nonumber\\
&=& \int  d\theta d\phi \, \frac{-\sin \theta \,
	w^\alpha}{2\left[ \left( z^\gamma(\tau+\epsilon)-z^\gamma \right)
	\, u_\gamma(\tau+\epsilon) \right]^2}\nonumber\\
&&\left( \left[ 1-u^\delta
u_\delta(\tau+\epsilon)+\left( z^\mu(\tau+\epsilon)-z^\mu \right) \, a_\mu(\tau+\epsilon)
\right] \right. \nonumber\\
&&\left( u^\beta+w^\beta \right) \, u_\beta(\tau+\epsilon)
-\left( z^\lambda(\tau+\epsilon)-z^\lambda \right) \,
u_\lambda(\tau+\epsilon) \, \nonumber \\
&&\left. \left[ \left( a^\kappa+\partial_\tau
w^\kappa \right) \, u_\kappa(\tau+\epsilon)
+\left( u^\nu+w^\nu \right) \, a_\nu(\tau+\epsilon) \right] \right) \nonumber\\
&=&\frac{-2\pi}{3 \left[ \left( z^\gamma(\tau+\epsilon)-z^\gamma
	\right) \, u_\gamma(\tau+\epsilon) \right]^2}\nonumber\\
&&\left( \vphantom{\Lambda^\gamma_\delta \, \Lambda^\alpha_\mu \,
	m^{\delta \mu}} \left[ 1-u^\delta
u_\delta(\tau+\epsilon)+ \left( z^\mu(\tau+\epsilon)-z^\mu
\right) \, a_\mu(\tau+\epsilon)
\right] \right. \nonumber\\
&&\hspace{1cm} u_\beta(\tau+\epsilon) \, \left( u^\alpha u^\beta -\eta^{\alpha \beta} \right) \nonumber \\
&&-\left( z^\lambda(\tau+\epsilon)-z^\lambda \right) \,
u_\lambda(\tau+\epsilon) \nonumber\\
&&\hspace{1cm} \left[ a_\nu(\tau+\epsilon) \, \left( u^\alpha u^\nu
-\eta^{\alpha \nu} \right) \right. \nonumber \\
&&\left. \left. \hspace{2cm}+u_\kappa(\tau+\epsilon) \, \partial_\tau
\Lambda^\kappa_\rho \, \Lambda^\alpha_\chi \, m^{\rho \chi} \right] \right) \, .
\end{eqnarray}
To simplify (\ref{long-int2}) further we need an expression for
$ \partial_\tau \Lambda^\kappa_\rho \, \Lambda^\alpha_\chi \, m^{\rho
	\chi}$. With the help of (\ref{m-rule}) for $m^{\alpha\beta}$ this
yields $a^\gamma u^\alpha$ for the term that contains the two
Kronecker deltas. To evaluate the term containing $\eta^{\alpha\beta}$
we go into the co-moving frame. The required Lorentz matrix is just a
unit matrix while its derivative contains only accelerations in the
time-space part as can be understood by considering the
non-relativistic limit. We find
\begin{eqnarray}
\partial_\tau \Lambda^\gamma_\delta \, \Lambda^\alpha_\mu \, \eta^{\delta \mu}&=&
\begin{pmatrix}
0&-a^1&-a^2&-a^3\\
a^1&0&0&0\\
a^2&0&0&0\\
a^3&0&0&0\\
\end{pmatrix} \nonumber \\
&=&a^\gamma u^\alpha-a^\alpha u^\gamma \, .
\end{eqnarray}
We note that in the rest frame the well-know Thomas
precession is absent. With both terms combined we get $ \partial_\tau
\Lambda^\gamma_\delta \, \Lambda^\alpha_\mu \, m^{\delta \mu}=u^\gamma
a^\alpha$. For integral $\fbox{iii}$ we obtain
\begin{eqnarray}
\fbox{iii}&=&\int d\theta d\phi \, \frac{\sin \theta\,u^\alpha}{2r}a^\beta
w_\beta \nonumber \\
&=&\int d\theta d\phi \, \frac{\sin \theta \,
	u^\alpha \left( u^\delta+w^\delta \right) \,
	u_\delta(\tau+\epsilon) \, a^\beta w_\beta }{2 \left(
	z^\gamma(\tau+\epsilon)-z^\gamma \right) \,
	u_\gamma(\tau+\epsilon)}\nonumber\\
&=&\frac{2\pi u^\alpha \, u_\delta(\tau+\epsilon) \, a_\beta \left( u^\beta u^\delta -\eta^{\beta
		\delta} \right)}{3 \left( z^\gamma(\tau+\epsilon)-z^\gamma
	\right) \, u_\gamma(\tau+\epsilon)}\nonumber\\
&=&\frac{-2\pi u^\alpha \, u_\delta(\tau+\epsilon) \, a^\delta}{3 \left(
	z^\gamma(\tau+\epsilon)-z^\gamma \right) \, u_\gamma(\tau+\epsilon)} \, ,
\end{eqnarray}
while integral $\fbox{iv}$ can be recast into
\begin{eqnarray}
\fbox{iv}&=&\int d\theta d\phi \, \frac{\sin \theta \,w^\alpha}{2r} \, a^\beta w_\beta 
\nonumber\\
&=&\int d\theta d\phi \, \frac{\sin \theta
	\,w^\alpha \left( u^\delta+w^\delta \right) \,
	u_\delta(\tau+\epsilon) \, a^\beta
	w_\beta}{2 \left( z^\gamma(\tau+\epsilon)-z^\gamma \right) \, u_\gamma(\tau+\epsilon)}\nonumber\\
&=&\frac{2\pi \, u^\delta u_\delta(\tau+\epsilon) \, a_\beta \left( u^\alpha
	u^\beta-\eta^{\alpha \beta}
	\right)}{3 \left( z^\gamma(\tau+\epsilon)-z^\gamma \right) \, u_\gamma(\tau+\epsilon)}\nonumber\\
&=&\frac{-2\pi \, a^\alpha \, u^\delta
	u_\delta(\tau+\epsilon)}{3 \left( z^\gamma(\tau+\epsilon)-z^\gamma
	\right) \, u_\gamma(\tau+\epsilon)} \, .
\end{eqnarray}
Integrals $\fbox{v}$ and $\fbox{via}$ give
\begin{eqnarray}
\fbox{v}&=&\int d\theta d\phi \, \frac{-\sin \theta \,w^\alpha}{2r^2}
\nonumber\\
&=&\int d\theta d\phi \, \frac{-\sin \theta
	\,w^\alpha \left[ \left( u^\beta+w^\beta \right) \,
	u_\beta(\tau+\epsilon)
	\right]^2}{2\left[ \left( z^\gamma(\tau+\epsilon)-z^\gamma \right)
	\, u_\gamma(\tau+\epsilon) \right]^2}\nonumber\\
&=&\frac{-4\pi \, u^\beta
	u_\beta(\tau+\epsilon) \, u_\delta(\tau+\epsilon) \, \left( u^\alpha u^\delta -\eta^{\alpha 
		\delta} \right)}{3 \left[ \left(
	z^\gamma(\tau+\epsilon)-z^\gamma \right) \,
	u_\gamma(\tau+\epsilon) \right]^2}
\end{eqnarray}
and
\begin{eqnarray}
\fbox{via}&=&\int    d\theta d\phi \, \frac{-u^\alpha}{2r^2} \, \partial_\phi
r \, a^\beta \phi_\beta \nonumber\\
&=&\int d\theta d\phi \, \frac{u^\alpha \sin \theta \, \phi^\delta \,
	u_\delta(\tau+\epsilon) \, a^\beta
	\phi_\beta}{2 \left( z^\gamma(\tau+\epsilon)-z^\gamma \right) \,
	u_\gamma(\tau+\epsilon)} \, .
\end{eqnarray}
Integral $\fbox{vib}$ can be evaluated similarly, only
$\phi^\alpha$ and $\theta^\alpha$ are exchanged. Integrals $\fbox{via}$ and
$\fbox{vib}$ yield together
\begin{eqnarray}
\label{via+vib}
&&\fbox{via}+\fbox{vib} \nonumber \\
&=&\frac{u^\alpha \, u_\delta(\tau+\epsilon) \, a_\beta}{2
	\left( z^\gamma(\tau+\epsilon)-z^\gamma \right) \,
	u_\gamma(\tau+\epsilon)} \nonumber\\
&& \left[ 4\pi \, \left( u^\beta u^\delta -\eta^{\beta
	\delta} \right)-\frac{4\pi}{3} \left( u^\beta u^\delta -\eta^{\beta
	\delta} \right) \right] \nonumber\\
&=&\frac{-4\pi \, u^\alpha \, u_\delta (\tau+\epsilon) \, a^\delta }{3
	\left( z^\gamma(\tau+\epsilon)-z^\gamma \right) \, u_\gamma (\tau+\epsilon)} \, .
\end{eqnarray}
The result (\ref{via+vib}) can be obtained by pulling a common angle
independent term in $\fbox{via}$ and $\fbox{vib}$ in front of the integrals. The
remaining term $\phi^\gamma \phi^\beta+\theta^\gamma \theta^\beta$ has
been replaced by $m^{\alpha \beta}-w^\alpha w^\beta$. After angle
integration essentially only $m^{\alpha \beta}$ remains, which can be evaluated to
$u^\alpha u^\beta-\eta^{\alpha\beta}$. The same situation is
encountered for all remaining integrals with labels a and b in
(\ref{long-eqn}). If we go through $\fbox{viia}+\fbox{viib}$ zero is obtained because
there are only odd powers of $w^\alpha$ in the expressions. The
remaining integrals are
\begin{eqnarray}
&&\fbox{viiia}+\fbox{viiib} \nonumber \\
&=&\int d\theta d\phi \, \frac{u^\alpha \sin \theta
	\,  u_\delta(\tau+\epsilon) \, \partial_\tau w_\beta
	\left( \phi^\beta \phi^\delta + \theta^\delta
	\theta^\beta \right)}{2 \left( z^\gamma(\tau+\epsilon)-z^\gamma
	\right) \, u_\gamma(\tau+\epsilon)} \nonumber\\
&=&0
\end{eqnarray}
and
\begin{eqnarray}
&&\fbox{ixa}+\fbox{ixb} \nonumber \\
&=&\int d\theta d\phi \, \frac{w^\alpha \sin \theta
	\, u_\chi(\tau+\epsilon) \, \partial_\tau w_\beta \, \left( \phi^\beta
	\phi^\chi+ \theta^\chi \theta^\beta \right)}{2 \left(
	z^\gamma(\tau+\epsilon)-z^\gamma \right) \,
	u_\gamma(\tau+\epsilon)} \nonumber\\
&=&\frac{u_\chi (\tau+\epsilon)
	\, \Lambda^\alpha_\sigma  \, \Lambda^\beta_\nu
	\, \Lambda^\chi_\xi  \, \eta_{\beta \delta} \, \partial_\tau
	\Lambda^\delta_\rho}{2 \left( z^\gamma(\tau+\epsilon)-z^\gamma \right) \,
	u_\gamma(\tau+\epsilon)} \, \left( \frac{4\pi}{3} m^{\sigma \rho}m^{\nu
	\xi} \right. \nonumber \\
&&\left. \hspace{0.5cm} -\frac{4\pi}{15} \left( m^{\sigma \rho}m^{\nu \xi}+m^{\sigma
	\nu}m^{\rho \xi}+m^{\sigma \xi}m^{\nu \rho} \right) \right) \nonumber\\
&=& \frac{u_\chi (\tau+\epsilon) \, \eta_{\beta\delta}}{2 \left(
	z^\gamma(\tau+\epsilon)-z^\gamma \right)
	u_\gamma(\tau+\epsilon)} \, \left( \frac{4\pi}{3} \, a^\alpha
u^\delta \left( u^\beta u^\chi-\eta^{\beta\chi} \right)
\right. \nonumber\\
&&-\frac{4\pi}{15} \, a^\alpha u^\delta \, \left( u^\beta
u^\chi-\eta^{\beta\chi} \right) +\left( u^\alpha
u^\beta-\eta^{\alpha\beta} \right) \, a^\chi u^\delta \nonumber \\
&&\left. \hspace{1cm}+\left( u^\alpha u^\chi-\eta^{\alpha\chi}
\right) \, a^\beta u^\delta \vphantom{\frac{4\pi}{15}} \right)=0
\end{eqnarray}
and
\begin{eqnarray}
&&\fbox{xa}+\fbox{xb} \nonumber \\
&=&\int  d\theta d\phi \, \frac{-\theta^\alpha \sin \theta \,\partial_\theta
	r-\phi^\alpha \partial_\phi r}{2r^3} \nonumber\\
&=&\int d\theta d\phi \,
\frac{ \sin \theta \, \left( \theta^\alpha \theta^\beta+\phi^\alpha
	\phi^\beta \right) \, u_\beta(\tau+\epsilon) \,
	u_\delta(\tau+\epsilon)}{2 \left[ \left(
	z^\gamma(\tau+\epsilon)-z^\gamma \right) \,
	u_\gamma(\tau+\epsilon) \right]^2} \nonumber \\
&&\hspace{1cm} \times \left( w^\delta+u^\delta \right) \nonumber\\
&=&\frac{4\pi \, u_\beta(\tau+\epsilon) \, u^\delta u_\delta(\tau+\epsilon) \left(
	u^\alpha u^\beta -\eta^{\alpha \beta} \right)}{3 \left[ \left( z^\gamma(\tau+\epsilon)-z^\gamma
	\right) \, u_\gamma(\tau+\epsilon) \right]^2}
\end{eqnarray}
and
\begin{eqnarray}
&&\fbox{xi}\nonumber \\
&=&\int     d\theta d\phi \, \frac{\sin \theta \, \left(
	a^\alpha+a^\delta w_\delta w^\alpha \right)}{r}
\nonumber\\
&=&\int d\theta d\phi \,\frac{\sin \theta \, \left( a^\alpha+a^\delta w_\delta
	w^\alpha \right) \, \left( u^\beta+w^\beta \right) 
	u_\beta(\tau+\epsilon) 
}{\left( z^\gamma(\tau+\epsilon)-z^\gamma \right)
\, u_\gamma(\tau+\epsilon)} \nonumber\\
&=&\frac{4\pi \, a^\alpha \, u^\beta u_\beta(\tau+\epsilon)}{\left(
	z^\gamma(\tau+\epsilon)-z^\gamma \right) \,
	u_\gamma(\tau+\epsilon)} \nonumber\\
&&+\frac{4\pi \, a_\delta \, u^\beta u_\beta(\tau+\epsilon) \, \left( u^\alpha u^\delta -\eta^{\alpha 
		\delta} \right)}{3 \left( z^\gamma(\tau+\epsilon)-z^\gamma \right)
	\, u_\gamma(\tau+\epsilon)} \nonumber\\
&=&\frac{8\pi \, a^\alpha \, u^\beta u_\beta(\tau+\epsilon)}{3\left(
	z^\gamma(\tau+\epsilon)-z^\gamma \right) \,
	u_\gamma(\tau+\epsilon)}
\end{eqnarray}
and
\begin{eqnarray}
&&\fbox{xiia}+\fbox{xiib}\nonumber \\
&=&\int d\theta d\phi \, \frac{u^\alpha+w^\alpha}{r^2} \left( a^\beta \theta_\beta \sin
\theta \,\partial_\theta r + a^\beta \phi_\beta \partial_\phi r \right) 
\nonumber\\
&=&\int d\theta d\phi \, \frac{-\sin \theta
	\, \left( u^\alpha+w^\alpha \right) \, a_\beta u_\delta(\tau+\epsilon) \, \left( \theta^\delta
	\theta^\beta+\phi^\delta \phi^\beta \right)}{\left( z^\gamma(\tau+\epsilon)-z^\gamma
	\right) \, u_\gamma(\tau+\epsilon)} \nonumber \\
&=&\frac{-8\pi \, u^\alpha \, a_\beta u_\delta(\tau+\epsilon) \left(
	u^\delta u^\beta -\eta^{\delta \beta} \right) }{3 \left(
	z^\gamma(\tau+\epsilon)-z^\gamma \right) \,
	u_\gamma(\tau+\epsilon)} \nonumber\\
&=&\frac{8\pi \, u^\alpha \, a^\delta u_\delta(\tau+\epsilon)}{3\left(
	z^\gamma(\tau+\epsilon)-z^\gamma \right) \, u_\gamma(\tau+\epsilon)}
\end{eqnarray}
and
\begin{eqnarray}
\label{larmor-formula}
&&\fbox{xiii} \nonumber \\
&=&\int d\theta d\phi \, \sin \theta \left(-\left( u^\alpha+w^\alpha
\right) \, \left[ a_\gamma a^\gamma+\left(
a_\gamma w^\gamma \right)^2 \right] \right) \nonumber\\
&=&-\frac{8\pi}{3}a_\gamma a^\gamma u^\alpha \, .
\end{eqnarray}
Not surprisingly (\ref{larmor-formula}), is the well know term contained in Larmor's formula. Now let us combine all terms. Equation
(\ref{long-eqn}), hence reads
\begin{eqnarray}
\label{pre-result-rad-react}
&&\int d\theta d\phi \,  n_\beta \, \frac{4\pi}{q^2} \,
T^{\alpha\beta} \nonumber\\
&=&\frac{2\pi}{3 \left[ \left( z^\gamma(\tau+\epsilon)-z^\gamma \right) \,
	u_\gamma(\tau+\epsilon) \right]^2} \nonumber\\
&&\times \left\{ \vphantom{u^\gamma u_\gamma(\tau+\epsilon) \left( u^\alpha
	u^\beta -\eta^{\alpha \beta} \right)}
-3u^\alpha \left[ 1-u^\delta
u_\delta(\tau+\epsilon)+\left( z^\rho(\tau+\epsilon)-z^\rho
\right) \right. \right. \nonumber\\
&&\left. \hspace{0.5cm} \times \, a_\rho(\tau+\epsilon) \right] \, u^\beta
u_\beta(\tau+\epsilon) \nonumber \\
&&+3u^\alpha \left( z^\mu(\tau+\epsilon)-z^\mu \right) \,
u_\mu(\tau+\epsilon) \, \nonumber \\
&&\hspace{0.5cm} \times \, \left[ a^\nu u_\nu(\tau+\epsilon)+u^\sigma
a_\sigma(\tau+\epsilon) \right] \nonumber\\
&& -\left[ 1-u^\chi u_\chi(\tau+\epsilon)+\left(
z^\xi(\tau+\epsilon)-z^\xi \right) \, a_\xi
(\tau+\epsilon) \right] \nonumber \\
&&\hspace{0.5cm} \times \, u_\lambda(\tau+\epsilon) \left( u^\alpha u^\lambda -\eta^{\alpha \lambda}
\right) \nonumber \\
&&+\left( z^\kappa(\tau+\epsilon)-z^\kappa \right) \,
u_\kappa(\tau+\epsilon) \nonumber \\
&&\hspace{0.5cm} \times \, \left[ a_\zeta(\tau+\epsilon) \left( u^\alpha u^\zeta 
-\eta^{\alpha \zeta} \right)+u_\sigma(\tau+\epsilon) \, u^\sigma
a^\alpha \right] \nonumber\\
&&-2u^\pi u_\pi(\tau+\epsilon) \, u_\psi(\tau+\epsilon) \left( u^\alpha
u^\psi -\eta^{\alpha \psi} \right)\nonumber\\
&&\left. +2u_\omega(\tau+\epsilon) \, u^\iota u_\iota(\tau+\epsilon) \left( u^\alpha
u^\omega -\eta^{\alpha \omega} \right) \right\} \nonumber\\
&&+
\frac{2\pi}{3 \left( z^\gamma(\tau+\epsilon)-z^\gamma \right) \,
	u_\gamma(\tau+\epsilon)} \nonumber\\
&&\times \left\{ -u^\alpha u_\eta(\tau+\epsilon) \, a^\eta -a^\alpha u^\upsilon
u_\upsilon(\tau+\epsilon) \right. \nonumber\\
&&-2u^\alpha u_\tau (\tau+\epsilon) \, a^\tau+4u^\alpha a^o
u_o(\tau+\epsilon) \nonumber\\
&&\left. +4 a^\alpha \, u^\vartheta u_\vartheta(\tau+\epsilon)
\right\} \nonumber \\
&&-\frac{8\pi}{3} \, a^\varphi a_\varphi u^\alpha \, .
\end{eqnarray}
After further simplification one arrives for the right-hand side of (\ref{pre-result-rad-react}) at
\begin{eqnarray}\label{onde}
&&\frac{2\pi}{3\left[ \left( z^\gamma(\tau+\epsilon)-z^\gamma \right) \,
	u_\gamma(\tau+\epsilon) \right]^2}\nonumber\\
&&\left\{ \left[ u^\alpha(\tau+\epsilon)-4u^\alpha u^\beta
u_\beta(\tau+\epsilon) \right] \right. \nonumber\\
&& \left. \times \left[ 1-u^\delta
u_\delta(\tau+\epsilon)+(z^\rho(\tau+\epsilon)-z^\rho) \,
a_\rho(\tau+\epsilon) \right] \right\} \nonumber\\
&&+
\frac{2\pi}{3 \left( z^\gamma(\tau+\epsilon)-z^\gamma \right) \, u_\gamma(\tau+\epsilon)}\nonumber\\
&&\left\{ 3u^\alpha \left[ a^\nu u_\nu(\tau+\epsilon)+u^\sigma
a_\sigma(\tau+\epsilon) \right] \right. \nonumber\\
&&+a_\zeta(\tau+\epsilon) \left( u^\alpha u^\zeta 
-\eta^{\alpha \zeta} \right)+u_\sigma(\tau+\epsilon) \, u^\sigma
a^\alpha \nonumber\\
&&\left. +u^\alpha u_\tau(\tau+\epsilon) a^\tau +3a^\alpha u^\vartheta
u_\vartheta(\tau+\epsilon) \right\} \nonumber \\
&&-\frac{8\pi}{3} a^\varphi a_\varphi u^\alpha
\nonumber\\
&=&\frac{2\pi}{3 \left[ \left( z^\gamma(\tau+\epsilon)-z^\gamma
	\right) \, u_\gamma(\tau+\epsilon) \right]^2}\nonumber\\
&& \left\{ \left[ u^\alpha(\tau+\epsilon)-4u^\alpha u^\beta
u_\beta(\tau+\epsilon) \right] \right. \nonumber\\
&& \left. \times \left[ 1-u^\delta
u_\delta(\tau+\epsilon)+(z^\rho(\tau+\epsilon)-z^\rho) \,
a_\rho(\tau+\epsilon) \right] \right\} \nonumber\\
&+&
\frac{2\pi}{3 \left( z^\gamma(\tau+\epsilon)-z^\gamma \right) \, u_\gamma(\tau+\epsilon)}\nonumber\\
&&\left\{ 4u^\alpha \left[ a^\tau u_\tau(\tau+\epsilon)+u^\zeta
a_\zeta(\tau+\epsilon) \right] \right. \nonumber \\
&&\left. +4a^\alpha u^\vartheta u_\vartheta(\tau+\epsilon) - a^\alpha(\tau+\epsilon) \right\} \nonumber \\
&&-\frac{8\pi}{3} a^\varphi a_\varphi u^\alpha \, .
\end{eqnarray}
This equation is not yet the final result. One step is still missing.
The effect of the time derivative and the time integral in
(\ref{totalint}) also have to be taken into account. Their combined
effect is the coordinate shift
$\tau\rightarrow\tau-\epsilon$. After reintroducing the factor
$-q^2/4\pi$, the full electromagnetic force is given by
\begin{eqnarray}
\partial_\tau P^\alpha_\epsilon (\tau)
= &&-\frac{q^2}{6 \left[ \left( z^\gamma-z^\gamma
	(\tau-\epsilon) \right) \, u_\gamma \right]^2}\nonumber\\
&&\left\{ \left[ u^\alpha-4u^\alpha(\tau-\epsilon) \, u^\beta
u_\beta(\tau-\epsilon) \right] \right.\nonumber \\
&&\left. \times \left[ 1 - u^\delta u_\delta(\tau-\epsilon)+\left( z^\rho-z^\rho
(\tau-\epsilon) \right) \, a_\rho \right] \right\} \nonumber\\
&&-
\frac{q^2}{6 \left( z^\gamma-z^\gamma(\tau-\epsilon) \right) \,
	u_\gamma} \nonumber \\
&&\left\{ 4u^\alpha
(\tau-\epsilon) \left[ a^\tau(\tau-\epsilon) \, u_\tau+u^\zeta
(\tau-\epsilon) \, a_\zeta \right] \right. \nonumber \\
&&\left. +4a^\alpha(\tau-\epsilon) \, u^\vartheta(\tau-\epsilon) \,
u_\vartheta-a^\alpha \right\} \nonumber\\
&& +\frac{2q^2}{3} \, a^\varphi(\tau-\epsilon) \, a_\varphi(\tau-\epsilon) \,
u^\alpha(\tau-\epsilon) \\
=: L^\alpha_\epsilon(\tau) &&\, .
\end{eqnarray}

\bibliography{radiation_reaction_finald.bbl}

\end{document}